\documentclass[11pt]{article}

\usepackage{sectsty}
\usepackage{graphicx}

\title{Socioeconomic reorganization of communication and mobility networks in response to external shocks}
\author{Ludovico Napoli$^1$, Vedran Sekara$^2$, Manuel Garc\'ia-Herranz$^3$, M\'arton Karsai$^{1,4}$\footnote{Corresponding author: \texttt{Karsaim@ceu.edu}}}
\date{}

\begin{document}
\maketitle	

% Optional TOC
% \tableofcontents
% \pagebreak

%--Paper--
\begin{center}
$^1$ Department of Network and Data Science, Central European University, Quellenstraße 51, 1100 Vienna (Austria) \\
$^2$ Department of Computer Science, IT University of Copenaghen, Rued Langgaards Vej 7, 2300 Copenhagen (Denmark) \\
$^3$ Frontier Data Tech Unit, Chief Data Office, UNICEF, 3 United Nations Plaza, New York, NY 10017 (United States) \\
$^4$ National Laboratory for Health Security, Alfréd Rényi Institute of Mathematics, Re\'altanoda utca, Budapest 1053, (Hungary) \\
\end{center}

\begin{center}
    \textbf{Abstract} \\
Socioeconomic segregation patterns in networks usually evolve gradually, yet they can change abruptly in response to external shocks. The recent COVID-19 pandemic and the subsequent government policies induced several interruptions in societies, potentially disadvantaging the socioeconomically most vulnerable groups. Using large-scale digital behavioral observations as a natural laboratory, here we analyze how lockdown interventions lead to the reorganization of socioeconomic segregation patterns simultaneously in communication and mobility networks in Sierra Leone. We find that while segregation in mobility clearly increased during lockdown, the social communication network reorganized into a less segregated configuration as compared to reference periods. Moreover, due to differences in adaption capacities, the effects of lockdown policies varied across socioeconomic groups, leading to different or even opposite segregation patterns between the lower and higher socioeconomic classes. Such secondary effects of interventions need to be considered for better and more equitable policies.
\end{center}

\pagebreak

\section{Introduction}
Segregation patterns between people~\cite{freeman1978segregation,reardon2011income,henry2011emergence} emerge through assortative biases that can appear along several individual characters. While gender, education, age or ethnicity are commonly identified dimensions along which segregation patterns evolve, studies showed that the socioeconomic status (SES) of people is another important determinant behind segregation~\cite{reardon2011income,musterd2017socioeconomic,leo2016socioeconomic}. People's SES, reflecting their wealth and social standing, is unevenly distributed in every modern society~\cite{kakwani1980income,piketty2014capital} and in turn induces inequalities across several contexts. It may stand behind the unequal access to better education~\cite{rumberger2005does}, occupation~\cite{hannah1989relationship} or to reach better health services and longer life expectancy~\cite{acevedo2003residential}. Similarly, SES largely determines people's friendship preferences, and thus affect both the local and global formation of their social networks~\cite{eagle2010network}. Indeed, social ties between people of different SES are not random, their emergence is governed by homophilic tie creation mechanisms~\cite{mcpherson2001birds}. Commonly, people tend to interact mostly with others of similar status, while scarcely building social ties with others from remote socioeconomic (SE) groups. The combined effects of status homophily and the uneven distribution of SES may induce assortative network configurations. They lead to observable segregation patterns where distinct SE groups are isolated from one another ~\cite{reardon2011income}, as it has been demonstrated earlier~\cite{iceland2006does, massey1987trends, musterd2017socioeconomic,schelling1971dynamic}.

Similar assortative patterns can be found in the mobility mixing of people. SES is a strong determinant of places people visit, the events they participate in, the transportation they use, and the time they travel. Due to these biases, people predominantly meet others coming from similar SE backgrounds, while they stay isolated from members of dissimilar SE groups. This induces observable segregation patterns in mobility mixing networks, as has been determined by several recent studies~\cite{dong2020segregated,xu2019quantifying,moro2021mobility,bokanyi2021universal,hilman2022socioeconomic}.

Segregation patterns in social and mobility networks are generally believed to be stable and only change gradually over years~\cite{farley1994changes,charles2003dynamics}. However, severe external shocks like dealing with a global pandemic might force people to change their behavior abruptly, leading to the sudden reorganization of their SE networks. During the early phase of the COVID-19 pandemic most countries introduced non-pharmaceutical interventions~\cite{perra2021non} aiming to reduce mobility and physical proximity of people to slow down outbreaks. While lockdowns and curfew periods were efficient to mitigate the number of new cases ~\cite{gozzi2021estimating, bonaccorsi2020economic, pullano2020evaluating, chinazzi2020effect, schlosser2020COVID}, their success came with high costs. They had severe short- and long-term consequences on several aspects of life ranging from economy and employment~\cite{green2021state}, to mental health issues~\cite{luchetti2020trajectory, alzueta2021COVID} and food consumption~\cite{alzueta2021COVID}. As interventions were directly aimed at reducing social contacts and mobility, they had unavoidable effects on SE networks as well due to differences in adaption capacity of different SE groups~\cite{pullano2020evaluating,gozzi2021estimating, duenas2021changes}. During interventions, wealthier people could adjust their mobility patterns easier by avoiding public transportation or working from home~\cite{pullano2020evaluating,lee2020human,hunter2021effect}. However, people of lower SES could not comply equally. Job insecurity and harder working conditions, that commonly required physical presence or were even considered ``critical", limited the capacity of lower SE groups to stay isolated~\cite{united2019world}. In turn, among other consequences such as limited access to resources or information, poorer people disproportionately suffered the burden of health consequences, being more exposed and reaching higher mortality during the pandemic~\cite{mena2021socioeconomic, levy2022neighborhood}.

\begin{figure*}[ht!]
  \includegraphics[width=\linewidth]{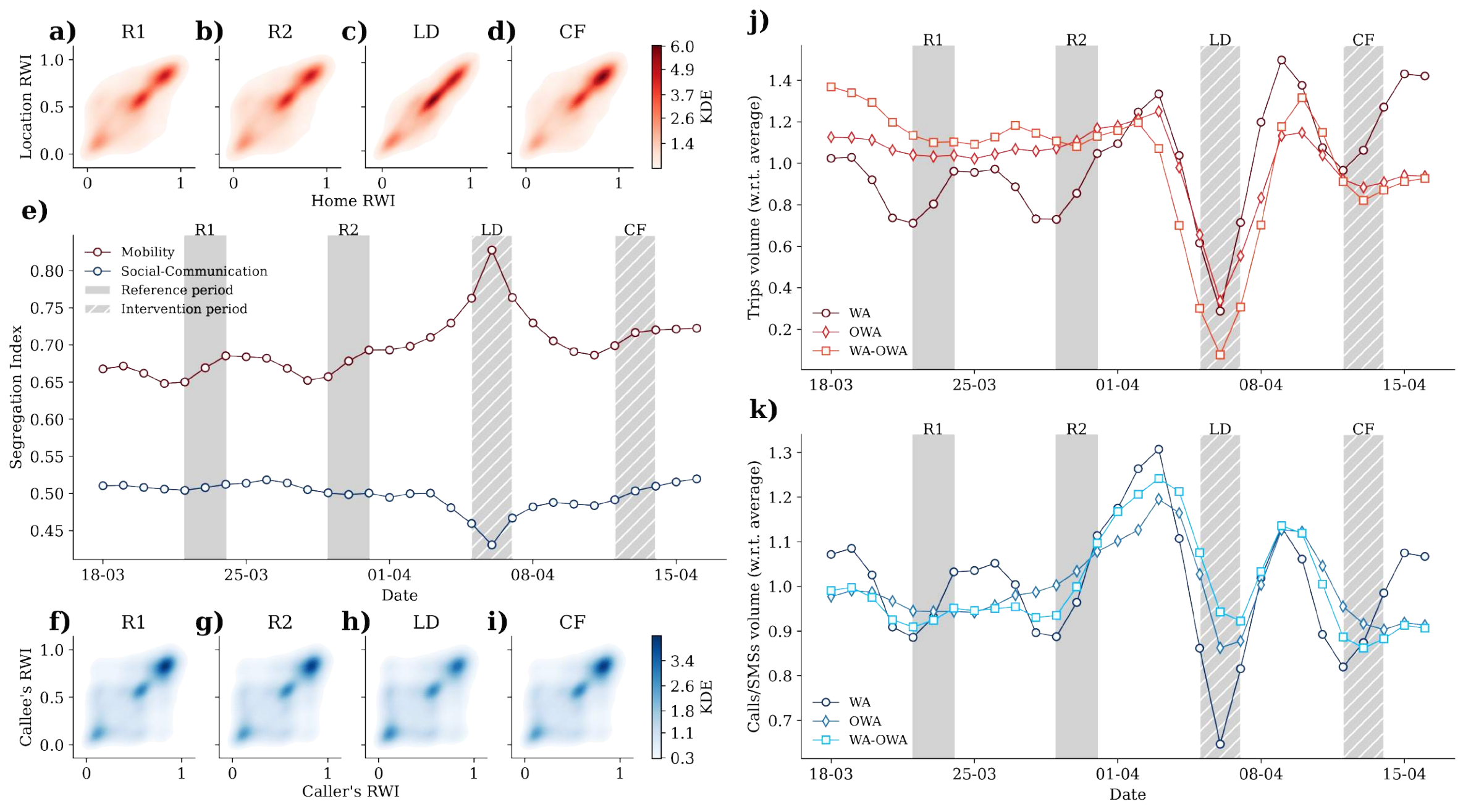}
\caption{\textbf{SE segregation dynamics in mobility and social communication networks.} (a-d) SE assortativity matrices (shown as the kernel density of the joint probability of RWIs) of the mobility network during the two reference (R1 and R2), lockdown (LD), and curfew (CF) periods. (e) The dynamics of the $\rho$ SE assortativity index computed for the mobility (red) and social communication (blue) networks. (f-i) Same as (a-d) but for the social communication network. (j) Relative number of travels within WA, OWA, and between the areas WA-OWA (also accounting for OWA-WA trips). (k) Number of communication events between people living in WA and OWA, or between the two geographic areas. All curves are normalized by their average computed over the full data period. For calculations on panels e, j and k we used 3-day symmetric rolling time windows with 1-day shift to obtain aggregated networks around the middle day at time $t$ of the actual window. For the corresponding non-aggregated results see SI Annex}
\label{fig1}

\end{figure*}

All these observations lead us to the question: \textit{How did these abrupt behavioral changes and differences in adjustment capacities re-organize the social and mobility networks of people in the short term}? To address this question we rely on a unique anonymized mobile phone call dataset collected in Sierra Leone during the beginning of the COVID-19 pandemic, that we combine with an inferred high-resolution SE map~\cite{chi2022microestimates}. Sierra Leone went through a series of episodes of containment measures during the pandemic: lockdowns, regional travel restrictions, and closing of places of worships. Our observation starts before the first pandemic interventions and includes the first national lockdown and curfew periods in April 2020. It provides us with the advantage to simultaneously follow the communication and mobility patterns of the same set of anonymized individuals. Analyzing these two behavioral aspects of approximately half million people, we follow the abrupt and short-term reorganization of their social and mobility networks and their altered SE segregation patterns due to external shocks, that leads us to the observation of potentially new socioeconomic phenomena.

\section*{Results}

\subsection*{Observation of network segregation patterns}

As an early response to the rapid global evolution of the COVID-19 pandemic in 2020, the government of Sierra Leone announced a three-day full lockdown, effective during April 5 to 7. That was followed by a 14 days nationwide nighttime curfew order starting from April 9, on top of other travel and shopping restrictions (more information can be found in the Supporting Information (SI Annex)). To understand how these interventions affected the mobility and communication behavior of people, we use data derived from a large anonymized and spatially aggregated Call Detailed Records (CDRs) data-sample (for details see Materials and Methods). Our observation starts on March 17, 2020, and spans over one month covering two weeks of reference period (which we call R1 and R2) and the restricted lockdown (LD) and curfew (CF) periods. Note that for fair comparisons these periods refer to the same weekdays (Sunday-Tuesday) during the observed weeks, corresponding to the weekdays of the LD period. To observe socioeconomic patterns, we first assign each tower location in our mobile phone data with an aggregate Relative Wealth Index (RWI) value \cite{chi2022microestimates}, then we infer users' home location~\cite{wesolowski2015quantifying, mari2017big} and assign them with the home location's associated RWI, which we use as a proxy to their SES (for details see Materials and Methods and SI Annex).
At the end of this pipeline, we obtain 405 unique locations with associated RWI and 505,676 individual mobile phone users with associated home location tower and RWI. Note that the outcome of our home location inference is strongly correlated with actual population density distribution both at the tower level and at all administrative levels (see SI Annex for more details), demonstrating that the mobile phone user population in focus is distributed spatially in a representative way in the country.

Using this coupled dataset, we build a social network ($G_S(t)$), with nodes as people and links as time-varying mobile communication interactions between them, and a mobility network ($G_M(t)$), where nodes represent the home and visited locations of people and temporal links capture their visiting patterns. Note, to avoid confounding spatial effects inducing spurious SE network correlations, we remove links that start and end in the same location (the main results without removals can be found in SI Annex). Nodes in the resulting networks, being people or places, are all associated with a corresponding RWI index (ranging between the poorest value 0 to the richest value 1), and segmented into nine equally populated SE classes, with the poorest labeled as 1 and the richest as 9 (for details see Materials and Methods).

SE network segregation patterns can be captured through the lens of network assortativity~\cite{newman2002assortative} that indicates connection preferences between similar nodes (being people or places). Segregation patterns can be visualised in the form of assortativity matrices~\cite{leo2016socioeconomic} that depicts the (communication or visiting) connection probabilities for people belonging to different SES. Finally, $\rho$ is the assortativity coefficient~\cite{newman2002assortative} (for definition see Materials and Methods), which summarizes the overall SE segregation captured by the assortativity matrix. It effectively indicates how much the assortativity matrix is concentrated around its diagonal. Like a correlation coefficient, $\rho$ is bounded between -1 (anti-segregation, also called disassortativity) and 1 (maximum segregation), with $\rho=0$ indicating no segregation in the network. Note that alternatively it is possible to use entropy-like metrics to quantify segregation in a population~\cite{eagle2010network, moro2021mobility}, but assortativity captures better network effects, like ones induced by homophilic mechanisms, captures better relative differences of SE diversity between egos and peers, and it overcomes data sparsity issues. (For a detailed analysis using entropy measures see SI Annex).

\begin{figure*}[ht!]
  \includegraphics[width=\linewidth]{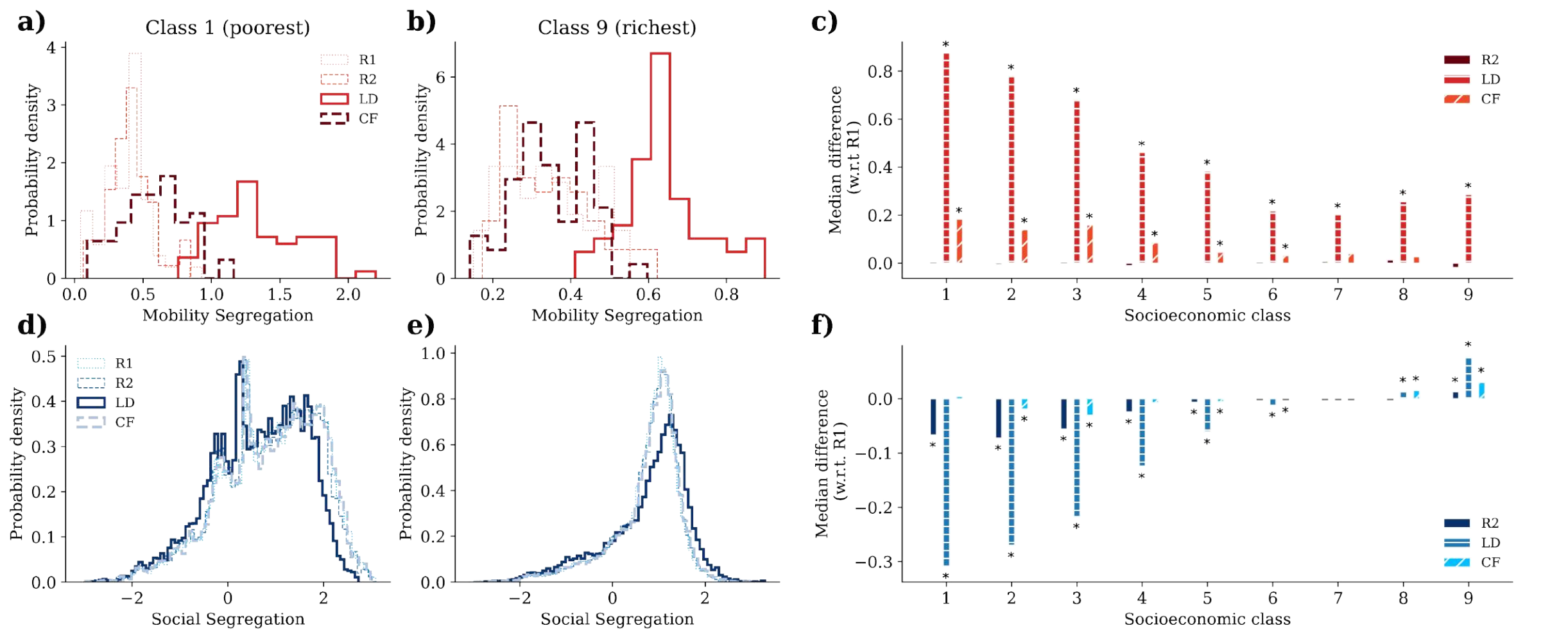}
  \caption{\textbf{Dynamics of individual-level segregation patterns.} The $P(r_u(t))$ individual assortativity index distributions computed from the mobility (panels a-b, in red) and social communication (panels d-e, in blue) networks for the poorest (class 1 in a and d) and the richest (class 9 in b and e) SE classes for the two reference periods (R1 and R2, thin dashed lines), the lockdown (LD, solid line), and curfew (CF, dashed thick line). Panels (c) and (f) depict the pairwise differences of median assortativity values of $P(r_u(t))$ for each nine SE group in the mobility and social networks (respectively). Differences are calculated between R1 and the R2, LD and CF periods. The asterisks symbols over the bars (when bars are positive, otherwise under them) in panels (c) and (f) indicate statistical significant differences computed with the one-tailed Mann-Withney U-test (with $\mbox{p-value} < 0.01$). The full list of p-values is shown in the SI Annex.}
\label{fig2}
\end{figure*}

\subsection*{Dynamics of network segregation patterns}

The lockdown had radical effects on the different activities of people, however, not to the same extent for each SE class. Both the average distance people traveled and the average number of calls/SMSs they made decreased during these days, but always with the richest people having the largest capacity to adjust to the emergency, as compared to their pre-lockdown behavior (see SI Annex). The assortativity matrix computed for the two pre-lockdown reference periods (R1 in Figure \ref{fig1}(a),(f) and R2 in Figure \ref{fig1}(b),(g)) appears with a strong diagonal component, demonstrating positive assortative mixing in the two networks. This is further confirmed by the relatively high assortativity indices, $\rho_M \sim 0.65$ for the mobility (red) and $\rho_S\sim 0.5$ for social communication (blue) curves in Figure \ref{fig1}(e) during R1 and R2 reference periods. In other words, during normal times both networks are organized in a highly segregated structure where people prefer to visit places or interact with peers from SE groups similar to their own, rather than mixing with other SE groups. Similar results have been found by Dong et al. ~\cite{dong2020segregated} with  $\rho\sim 0.4-0.8$ both in mobility and online networks in multiple countries. Note that we identified several confounding effects that could lead to similar network segregation patterns (e.g. local visits or interactions, business hour activities, urban/rural factors, or physical distance), however, even when we account for them systematically, our observations remain robust (See SI Annex).

These segregation patterns radically changed during the national lockdown that started on the April 5, 2020 (LD period in Figure \ref{fig1}(e)). Once the lockdown was announced (April 1, 2020), the mobility segregation index started to increase and reached its maximum during the days of the lockdown. This significant increase of segregation in mobility was expected, and has been observed in other studies ~\cite{bonaccorsi2020economic,glodeanu2021social}. During the lockdown, non-essential workplaces were closed and a stay-at-home order was put in place. This greatly restricted people's mobility mixing as their movements were concentrated around their residential places. This is also reflected by the stronger diagonal component of the mobility assortativity matrix in Figure \ref{fig1}(c) as compared to the reference periods (Figure \ref{fig1}(a),(b)). Interestingly, the increased segregation patterns were not persistent after the lockdown and relaxed back close to their pre-lockdown level during the subsequent curfew (CF) period (Figure \ref{fig1}(d),(e)). This is somewhat in contrast with other observations, where stronger mobility segregation patterns were found to be residual in US cities after lockdown periods as compared to pre-COVID reference values~\cite{yabe2023behavioral}.

Communication events do not have the same spatial constraints as mobility, thus their reorganization induced a fundamentally different segregation dynamic in the social network (see Figure \ref{fig1}(e)). Strikingly, we found that SE assortativity in the social network decreased and reached a minimum during the lockdown. The social network reorganized into a less segregated configuration, where communication between different SE classes increased as compared to the pre-lockdown periods (R1 and R2 in Figure \ref{fig1}(e)). The opposite segregation trends in the mobility and social behaviors may suggest that with people's mobility restricted, they compensated for the lack of physical contacts by communicating with peers from other SE classes. Interestingly, similar to mobility, the altered segregation patterns in the social communication network did not last long and soon relaxed back to their pre-lockdown level. Note that all results reported in Figure \ref{fig1} show average values aggregated in a 3-day long sliding time-window with 1 day shift. For the corresponding non-aggregated results see the SI Annex.

\subsection*{SE network reorganization}

The baseline segregation levels and the strikingly different network reorganization patterns are both rooted in the SE structure and the strong urban-rural division of Sierra Leone. There is a disproportionally higher concentration of wealthy people in the capital, Freetown, which lies in the largely urban administrative province \textit{Western Area} (WA), than in the rest of the country (\textit{outside of the Western Area} (OWA)), which is considerably more rural and has, on average, two times higher multidimensional poverty rates~\cite{Leone2019Sierra} (see the SI Annex for the precise geographical division of WA and OWA). While local spatial effects do not explain alone the observed segregation patterns (see SI Annex), interactions (either movements or communications) between WA and OWA play important roles in shaping the overall network segregation. To demonstrate this, we assign each edge of the $G_M$ mobility (or $G_S$ social) network into three classes, based on whether both (WA), none (OWA), or one node (WA-OWA) is located in the Western Area. Note that mobility links always represent movements from people's home to out-of-home locations, while social links are always between pairs of people with different inferred home location.

To observe network reorganization we follow the relative changes in the mobility and communication volumes over time, by measuring the number of interactions within or between the WA and OWA areas, relative to the their average calculated over the whole period. We find that the relative mobility (see Figure \ref{fig1}(j)) changed abruptly during lockdown. The number of trips radically decreased prior to the days of intervention (likely an effect of the early announcement of the restriction). During lockdown the largest relative decrease ($\sim 95\%$) is observed for travels between WA-OWA, while trips inside WA and OWA show a lesser relative decrease compared to the reference periods. This disappearance of long-distance travels between urban and rural areas largely contributes to the increase of segregation in the mobility network during lockdown as it amplifies the relative contribution of trips between nearby census areas with similar SE values (i.e. within WA and OWA). 

We find opposite dynamics in the reorganization of the social communication network (Figure \ref{fig1}(k)). While the communication volume first increases prior to the lockdown (indicating a form of coordination), it radically decreases later, right before the intervention period, across all categories. Nonetheless, interactions within WA decreases the most ($\sim 30\%$); the decrease within OWA is smaller ($\sim 10\%$), but the relative communication volume between WA and OWA does not change considerably during the days of lockdown. This larger reduction of communication within each area and the relative increase of importance of communication between the richer WA and the poorer OWA areas account for the overall decrease in segregation that we observed in Figure \ref{fig1}(e). Consequently, during lockdown, the wealthiest (WA) and poorest areas (OWA) of the country became relatively less connected both in terms of physical movements and in social communication. However, while in terms of mobility, segregation increases globally as travels were restricted to local trips only, in terms of social communication, the elevated relative importance of long-distance communications between urban and rural areas induces a global decrease of network segregation.

\begin{figure*}[ht!]
  \includegraphics[width=\linewidth]{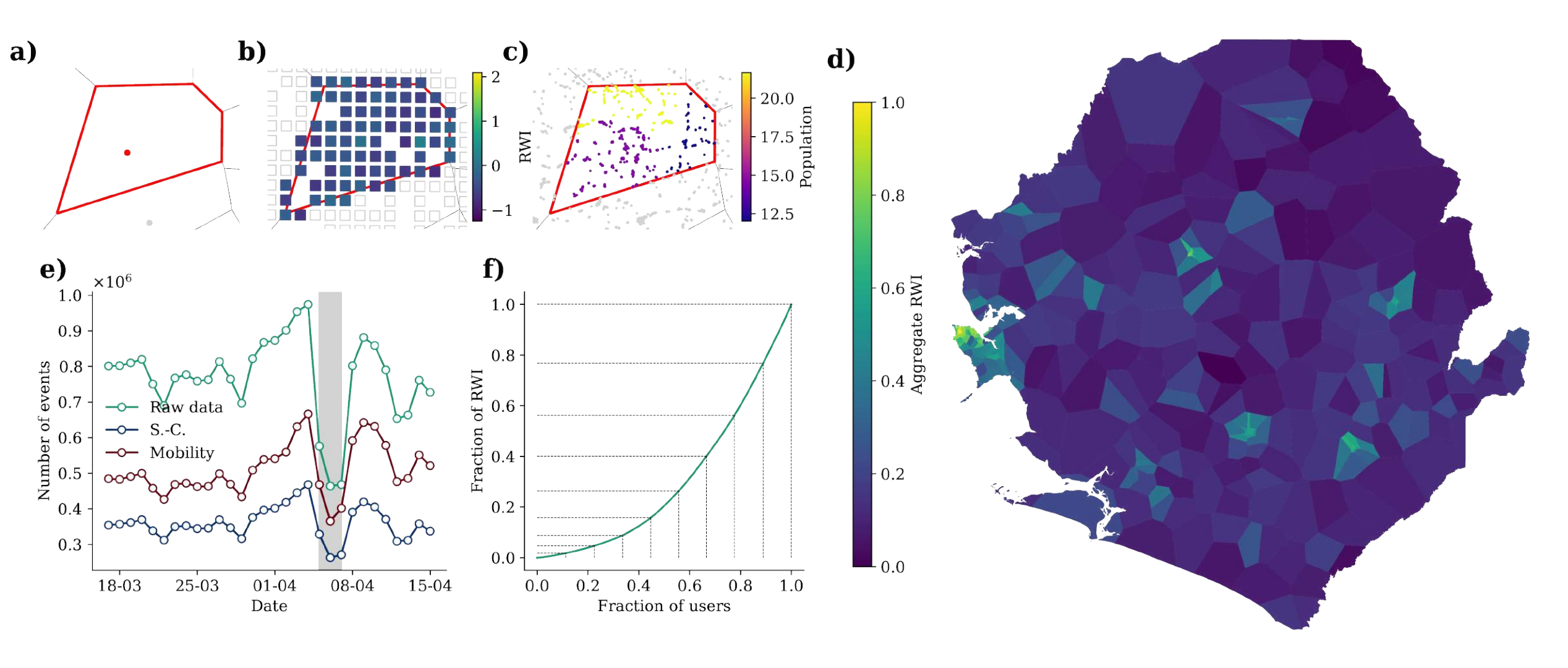}
  \caption{\textbf{Data combination pipeline.} (a) Single tower location and the respective Voronoi cell; (b) RWI patches overlapping with the corresponding cell; (c) population density distribution inside the same cell. Population density in each cell is used to weigh the RWI patches at the aggregate level. (d) Aggregate SE map of the whole Sierra Leone obtained from the aggregation procedure (panels a-c), after the normalization of the aggregated RWI values between 0 and 1. (c) Dynamics of the recorded data volume measured as the daily number of data points in the raw data (green curve), in the social-communication network (blue curve), and in the mobility network (red curve). Shaded area refers to the lockdown intervention period. (f) SE class segmentation based on the empirical Lorenz curve defined as the cumulative fraction of RWI of the sorted fraction of individuals by their inferred SES.}
  \label{fig3}
\end{figure*}

\subsection*{Individual-level segregation}

So far we have studied the segregation dynamics at a global network level but we learned less about individual behavioral responses. Beyond observing how SE segregation changed due to the reorganization of personal networks, this way of analysis can show whether external shocks affected different SE classes differently. To follow segregation changes of an individual $u$ in the network (a location in the mobility or a person in the social network), we compute the individual assortativity index, $r_u(t)$, that was introduced by Peel et al. in ~\cite{peel2018multiscale} (for definition see Materials and Methods). This measure quantifies the homogeneity of the local network around a central node $u$ in terms of the SES of neighbors as compared to the status of the central node. Computing $r_u(t)$ for each node in a SE class, the segregation dynamics of a SE class can be followed through the changes in the corresponding $P(r_u(t))$ individual assortativity index distribution. Note that $r_u(t)$ is an unbounded metric, which can take both positive and negative values, denoting assortative and disassortative mixing (respectively).

For mobility, people from both the lowest (class 1, Figure \ref{fig2}(a)) and highest SE classes (class 9, Figure \ref{fig2}(b)) exhibit positive assortativity values during reference periods, suggesting a baseline mobility segregation to be present in any circumstances, although with considerable variability. The lowest class shows stronger mobility segregation with a median value around $0.4$, as compared to the highest class with median $0.32$ (for precise values of all classes see Table S1 in SI Annex). The assortativity values do not change significantly between the reference periods (R1 and R2 in Figure \ref{fig2}(a),(b)). However, the lockdown introduces a significant upward shift in the assortativity distributions of each class, but with varying magnitude among classes (the median differences from the R1 reference distribution can be seen in Figure \ref{fig2}(c), LD bar). More precisely, the lockdown induces more than three times larger median assortativity values (changing from $\sim 0.41$ to $1.28$) for the poorest class while it almost doubles it for the richest one (from $\sim 0.34$ to $0.62$). Consequently, people from lower SE classes became more segregated in terms of mobility, while individuals from middle and higher SE classes got segregated, but to a lesser degree (one-tailed Mann-Witmann U Tests compared to reference period R1 indicates that all differences are significant). During the post-lockdown curfew (CF) period, mobility segregation patterns relaxed back close to their reference state for the wealthiest SE groups, however, the lower SE classes experienced some residual segregation after the lockdown (see CF bars in Figure \ref{fig2}(c)).

Similar to mobility, during the reference period the individual assortativity indices in the social communication network were mostly positive. They appear with smaller median values of $1.18$ for the lowest class and indicate stronger segregation for the richest class with a median around $1.28$ (see Figure \ref{fig2}(d),(e) and Table S1 in SI Annex). While this demonstrates a baseline positive assortativity in the social communication network too, $P(r_u(t))$ appears with negative values at the lower extreme, signaling some individuals with disassortative social mixing patterns.

The lockdown reorganized the social communication network in an unexpected way. While the global network assortativity already indicates an overall decrease in network segregation, the individual level observations shows that this was not homogeneous among the different SE classes. As shown in Figure \ref{fig2}(d), the main drivers behind the global decrease of assortativity are individuals from lower SE classes. For the poorest class, assortativity was reduced from the reference period median of $1.24$ to $1.02$ during LD (see Table S1 in SI Annex for all values). This appears in Figure \ref{fig2}(f) as negative median differences (LD bar) as compared to the R1 values. Thus, during this period, poorer people get in touch with a more diverse set of peers from higher SE classes, shifting their $P(r_u(t))$ to the left (in Figure \ref{fig2}d). Meanwhile, the wealthier classes got more segregated during LD. This is evident from the right-shifted LD distribution in Figure \ref{fig2}(e) and from the increased median assortativity ($1.38$) as compared to the reference period R1 value ($1.28$), leading to positive median differences in Figure \ref{fig2}(f). As a result, due to the dominant number of people from poorer classes with largely decreased individual network assortativity, the global network assortativity index (in Figure \ref{fig1}(c)) moderately decreases. Interestingly, contrary to mobility, here the top SE classes display some positive residual assortativity during the CF period (in Figure \ref{fig2}(e)), as they remain isolated even after LD. Note that spatial effects are not sufficient to explain the observed level of assortativities neither in the mobility nor in the social communication networks. For the corresponding analysis see SI Annex.

All these results clearly demonstrate that interventions could have very different secondary effects on communities and people of different SE background. Focusing exclusively on mobility or social communication will only reveal a part of the picture of how people adapt their behavior during emergencies. Our analysis underlines that SE background of people is an important determinant of behavioral response to external shocks, which can only be fully untangled by simultaneously following multiple aspects of human behavior.

\section*{Discussion}

Similar to other forms of segregation, SE assortative mixing patterns in mobility and social structures are assumed to change only slowly, conserving the status quo of the social divide. However, in this paper we empirically demonstrate that SE networks can change abruptly due to external shocks, and exhibit segregation patterns that affect people differently depending on their SE background on the short run. We found significant and opposite trends in the mobility and communication network segregation dynamics, changing due to the COVID-19 lockdown in Sierra Leone. We argue that these are due to the presence of different physical constraints, allowing different long-distance reorganization of these SE networks. At the individual level we detect significant differences in the segregation change of people in different SE classes. Although mobility networks become more segregated during lockdowns for everybody, due to the constrained movement of people between places, we find the segregation increase to be significantly more prominent and lasting longer for poorer SE classes. Interestingly, for social networks we find opposite effects, with segregation decreasing for poorer classes people, while increasing for richer ones.

Our study focuses on spatially embedded attributed networks, where the observed correlations could emerge due to certain confounding factors. To eliminate these effects we demonstrate that neither distance, nor confounding behavioral or network characters can explain the observed phenomena (see SI Annex), leaving us with the conclusion that they were induced by meaningful SE network patterns. Nevertheless, our results have certain limitations in terms of generalization, as they are based on observations of a single country. While the simultaneous observation of multiple behavioural aspects (like mobility and communication) of the same population over the same crisis period is difficult, we hope that future studies will able to meet this challenge to generalise further our results.  In terms of data quality, we analyze a SE map with varying spatial resolution determined by the location of mobile cell towers, that results occasionally in very large SE tracts at rural areas. Moreover, mobile phones in Sierra Leone may not be always personal devices but could be used by multiple people~\cite{erikson2018cell}, which could appear as noise in our observations. Finally, the observation period was limited to 4 weeks, making it impossible to observe longer-term residual effects of external shocks on SE network segregation. This limited observation time window, however, has been sufficient to assess abrupt behavioural changes due to restriction policies and their short-term consequences (for further discussion see SI Annex).

In summary, these results demonstrate the importance of socioeconomically disaggregated analyses of human behavior following simultaneously multiple behavioral traits. Beyond scientific merit, a better understanding of socioeconomically stratified mobility and social interaction patterns is important for epidemic modeling and policy design. This can help to avoid the uneven burden of direct and secondary impacts of interventions on economically disadvantaged populations~\cite{tizzoni2022addressing}. Following the reorganization of social-communication networks can also contribute to a better understanding of hidden dependency patterns in socioeconomically diverse societies. This can contribute to a better design of information dissemination campaigns and to better understand needs and emerging vulnerabilities.

\subsection*{Data description}

Results in this paper are based on the analysis of anonymized mobile phone Call Detailed Records (CDRs) provided by a major mobile phone operator in Sierra Leone that cover the communication activity of 1,270,214 anonymized users (16\% of population) between 17th March and 17th April 2020. Each data point in the CDRs contains information on a single communication event: the anonymized  IDs of the caller and the callee, the event's date, time and type (call or SMS), and the ID of the cell tower handling the event on the caller's side. The anonymization of customer IDs were done by random hashing by the provider, while spatial information about users' whereabouts was shared only at the granularity level of cell towers, making it impossible to infer the customer's identity and precise location.

\subsection*{Data processing}

Spatial information is crucial for us to track mobility movements of individual users at the cell tower level. By grouping together tower IDs with the same spatial coordinates (less than 1 meter of distance), we identified 405 unique tower locations. We assign to each tower a SE indicator by aggregating recently released, fine-grained estimates of Relative Wealth Index~\cite{chi2022microestimates} (RWI), that cover the area of Sierra Leone with 8,435 square patches with 2.4km side (see Figure \ref{fig3}(b)). To aggregate the RWI at the level of a cell tower, we consider for each of them the spatial intersection with the area covered by the tower, that we estimate to be the associated Voronoi cell (see Figure \ref{fig3}(a)), i.e. the area that includes all locations that are closer to that specific tower than to any other~\cite{aurenhammer2000voronoi}. Secondly, we assign to each patch a weight, defined by the population (obtained from a high-resolution population density map \cite{dataforgood}) that lives in the above-mentioned intersection between the patch and the Voronoi cell (see Figure \ref{fig3}(c)). Lastly, we use these weights to compute a weighted average of the RWIs of all patches that intersect with the cell, and this will be the SE indicator associated to the tower, that we keep calling RWI for simplicity.  Note that we scaled all the aggregate RWI values between 0-poorest and 1-richest for simplicity. A schematic representation of the aggregation procedure and the resulting SE map are shown in Figure \ref{fig3}(a),(b),(c),(d). As can be seen from the Figure \ref{fig3}(d), the resulting cells in the map have heterogeneous sizes that come from the uneven spatial distribution of towers, being mostly concentrated around the capital city Freetown and the other major cities; consequently, the spatial resolution of our data is necessarily higher in urban areas than in rural ones.

\subsection*{Home location and SES inference}

To infer a user's SES we couple the communication dataset with the SE map~\cite{chi2022microestimates}. We consider the spatial coordinates of the tower handling a communication event as a proxy to the caller's current location. We define a user's home location as the tower where most of the user's activity has been recorded during night time (9PM-6AM), lockdown, and curfew periods (see SI Annex for details). We assign the RWI of users' respective home location as their SES. After some filtering techniques (see SI Annex) we get a sample of 505,676 users with inferred SES. The Lorenz curve from our sample is shown in Figure \ref{fig3}(f). We get a Gini coefficient of 0.38, which is comparable to the World Bank value of 0.36~\cite{worldbankgini}. We group mobile phone users (and their home locations) in nine equally populated SE classes, shown in Figure \ref{fig3}(f) (see SI Annex for details). We also assign each tower location with a SE class, which is the same of users who were assigned that tower as inferred home location.

\subsection*{Mobility and social-communication network}

The mobility network $G_M(t)$ is constructed by mapping all communication events (\textit{caller,callee,tower}) into mobility events (\textit{home tower, tower}), where $home tower$ refers to the caller's inferred home location and $tower$ is the location where the caller was located at the time of the event. Therefore, nodes in $G_M(t)$ are tower locations, and events are home-to-location movement events. Given the restricted events for which we have the tower ID information and the restricted number of towers for which we infer the aggregate RWI, the number of movement events is also restricted as can be seen in Figure \ref{fig3}(e).

In the social-communication network $G_S(t)$, each node is a distinct mobile phone user and link weights are the number of communication events observed between pairs of nodes. We only consider events (\textit{caller, callee}) for which both the caller and the callee have an inferred home location and thus an inferred RWI value. For this reason, the effective number of communication events is also lower than the number of events in the raw data as can also be seen in Figure \ref{fig3}(e).

\subsection*{Global segregation}

The assortativity coefficient \cite{newman2002assortative} that we use to measure segregation in the global network is obtained by mapping all the CDR events $(i,j)_t$ observed at time $t$ to respective events $(s_i, s_j)_t$, where $i$ and $j$ are nodes in either $G_S(t)$ or $G_M(t)$ and $s_i$ and $s_j$ are their respective RWI values. The assortativity coefficient $\rho(t)$ is computed as the Pearson correlation between the $s_i$ and $s_j$ SE label of the interacting node pairs at time $t$. 

\subsection*{Individual segregation}

We define individual segregation index of node $u$ at time $t$ as:
$$
r_u(t) = \sum_{ij} w_{multi} (i;u) \frac{A_{ij}(t)}{k_i(t)}\tilde{x_i} \tilde{x_j},
$$
where $w_{multi}(i;u)$ is the multiscale distribution defined in \cite{peel2018multiscale}, $A_{ij}(t)$ is the adjacency matrix of either $G_S(t)$ or $G_M(t)$, $k_i(t)$ is the degree of node $i$ at time $t$, and $\tilde{x_i} = (x_i - \tilde{x}) / \sigma$ is the standardized SES. The measure describes how the presence of links in the local neighborhood of a single nodes co-varies with the properties (here SES) of the neighboring nodes. In our case, a positive individual assortativity indicates the tendency for a node to connect to others with similar SES. A negative value indicates the tendency of the central node to connect to other nodes with dissimilar SES. For scalar attributes (as in our case), the measure is not bounded: the higher the (absolute) value, the higher the assortativity (either positive or negative). For details and the complete list of values see SI Annex.

\subsection*{Acknoledgements}

The authors want to thank the Directorate of Science, Technology and Innovation (DSTI) and acknowledge their mandate to use science, technology, and innovation to support the Government of Sierra Leone to deliver on its national development plan effectively and efficiently. The authors are thankful for L. Peel for common scientific discussions. UNICEF is grateful for Takeda for their Investment in Innovation and Frontier Technology for better health through UNICEF Venture Fund and MagicBox initiatives. The authors are thankful for the collaboration of UNICEF Sierra Leone, and Shane O’Connor and for their field support and interest in Big Data and Complex Systems Science for Social Good. The authors are grateful for Victor Lagutov for his insightful comments on the manuscript.

\section*{Data and code availability}
Raw data is not publicly available due to privacy considerations and was made available for the specific purpose of this research, according to best practices of data protection. UNICEF's Frontier Data Technology Unit can individually assess specific requests to access the data with the intent of validation or furthering the value of this line of analysis in order to facilitate access to the data of this study.

\section*{Funding}

MK acknowledges support from the DataRedux ANR project (ANR-19-CE46-0008), the SoBigData++ H2020 project (SoBigData++ H2020-871042), the SAI Horizon 2020/CHIST-ERA project that was supported by FWF (I 5205-N), the EmoMap CIVICA project, and the National Laboratory of Health Security (RRF-2.3.1-21-2022-00006).

\newpage

\begin{center}
    \LARGE Supplementary Information \\

\end{center}

\section*{Spatial mapping}

Call Detail Records (CDRs) are mobile phone data providing details about communication events between users (calls or SMSs), such that each data point comes with the following (and more) information:

\begin{itemize}
    \item anonymized caller ID
    \item anonymized callee ID
    \item date and time 
    \item event type (call or SMS)
    \item cell tower ID
\end{itemize}

\begin{figure}[ht!]
    \centering
  \includegraphics[width=1\linewidth]{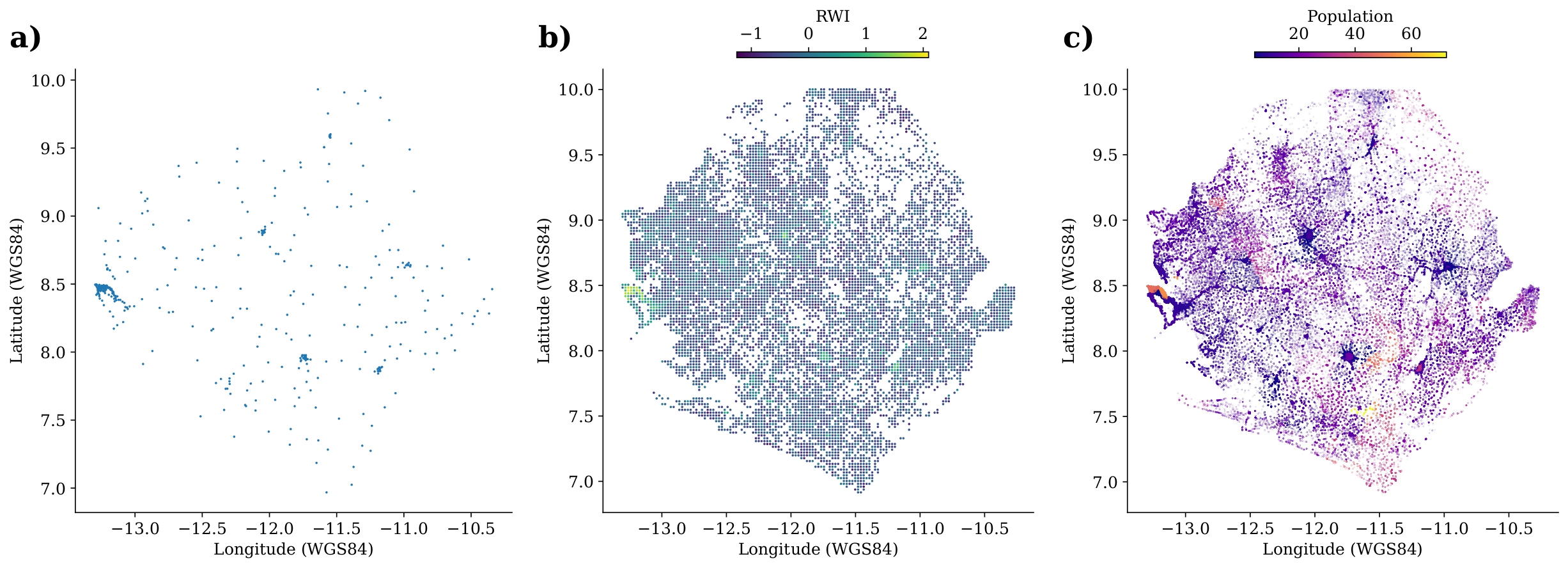}
  \caption{\textbf{Raw geospatial data.} (a) Spatial distribution of cell towers. (b) Spatial distribution of RWI values. (c) Spatial distribution of population. All the data shown here (raw data) are points georeferenced in WGS84 (longitude and latitude). }
  \label{figs7}
\end{figure}

The cell tower ID is a numerical identifier of the mobile company tower that handled a given communication event. Its geographical coordinates can be used as a proxy for the caller's current location at the time of communication. We use this information to infer the home location of mobile phone users, such that each user is assigned a cell tower as their home location (more details on this can be found in the next section). Cell tower locations are points and are referenced in geographic coordinates (longitude and latitude, WGS84) and their spatial distribution can be seen in Fig. \ref{figs7}(a).

We need to assign each cell tower location with a socioeconomic value regarding its level of wealth. We use the microestimates obtained from ~\cite{chi2022microestimates}, which are also points referenced in geographical coordinates (WGS84). However, they are built on a regular square grid with a 2.4 km side in the WGS84/Pseudo-Mercator projection. This means that each RWI value refers to people living in a 2.4 km-sided square cell on a planar surface. The data (points in WGS84) are shown in Fig. \ref{figs7}(b) and are color-coded according to the RWI value. 

We want to spatially aggregate this dataset such that each cell tower location is assigned one unique RWI value. In doing so, we need to consider the population density distribution. We use the high-resolution population density map obtained from Meta ~\cite{dataforgood}. These population data are also points in geographical coordinates WGS84 but refer to a 1 arcsec grid, which means that each data point estimates the number of people living in a 1 arcsec-sided cell on a spherical surface. The data is shown in Fig. \ref{figs7}(c).

To recap, we have three different georeferenced datasets:

\begin{enumerate}
    \item Cell tower locations: points in WGS84
    \item RWI cells: points in WGS84 referring to 2.4 km - side cells in the WGS84/Pseudo-Mercator projection
    \item Population density: points in WGS84 referring to 1 arcsec - side cells.
\end{enumerate}

\begin{figure}[ht!]
    \centering
  \includegraphics[width=1\linewidth]{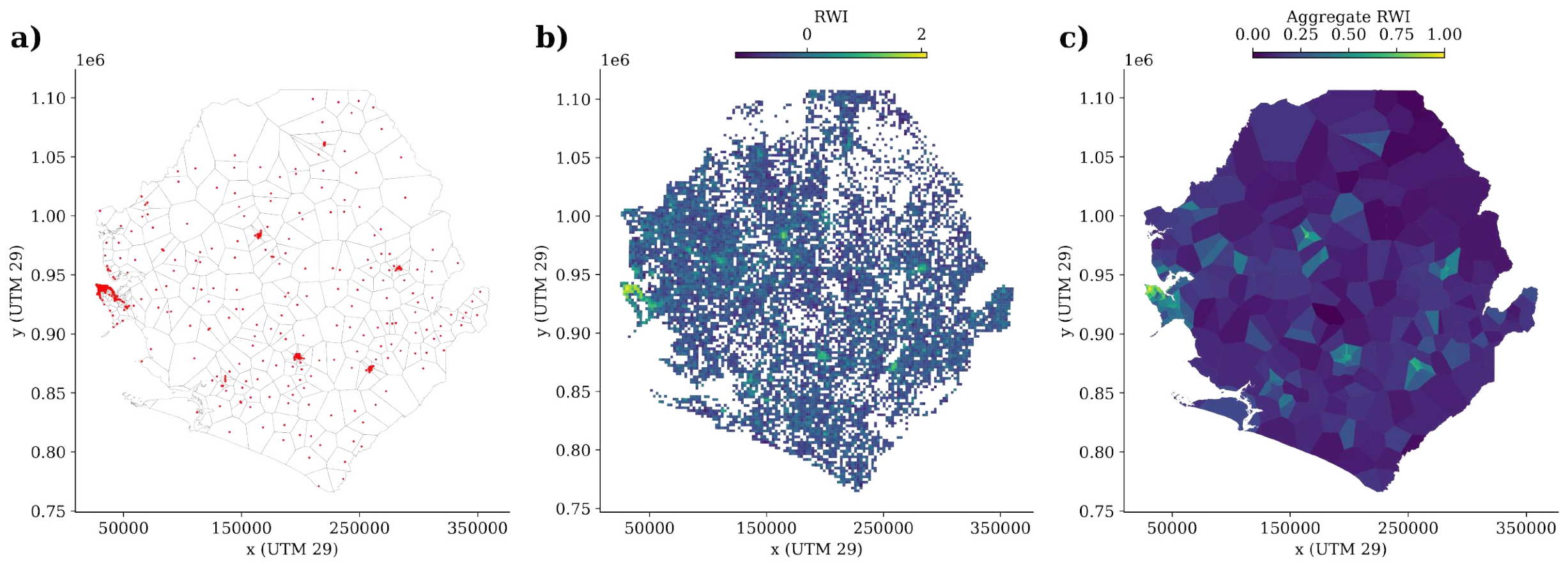}
\caption{\textbf{Processed geospatial data.} (a) Voronoi tessellation (grey polygons) obtained from cell towers (red points). (b) RWI square cells. (c) Aggregated RWI map. All the data shown here are polygons georeferenced in UTM 29.  }
  \label{figs8}
\end{figure}

We assume that the area covered by cell towers is best approximated by the Voronoi tessellation. The idea behind the Voronoi tessellation is that the border between two adjacent areas lies at exactly halfway between the two respective cell towers. The result is that a given tower's Voronoi cell contains all and only the locations that are closer to the given tower than to any other \cite{aurenhammer2000voronoi}. To obtain this, we first project the cell tower points to the projected system UTM 29 (where approximately half of Sierra Leone lies) and then use Euclidean distance to run the algorithm for Voronoi tessellation. In the end, for every Voronoi cell, we take the intersection with the borders of Sierra Leone, which is taken from the census and projected to UTM 29. The result is shown in Fig. \ref{figs8}(a).

Now we need to aggregate the RWI values over the Voronoi cells, taking into consideration the population density. First, we project the RWI points to WGS84/Pseudo-Mercator (where the squared cells are defined). In this projected system, we build the cell grid: for each point, we build a square polygon with a 2.4 km side such that the point will be in the middle of the polygon. We then project this polygon dataset to the same projected coordinate system of the Voronoi tessellation (UTM 29). The result is shown in Fig. \ref{figs8}(b). 

Regarding the population map, even though the dataset is defined as a 1 arcsec grid, we do not build the polygon dataset and project it, but we simply project the points to the UTM 29 coordinate system and discard the areas.

Given the extremely high resolution of the population data (one arcsec is around 30 meters, and the maximum value taken by a single population cell is 73 people) we do not think it's necessary to consider the spatial extension of population cell grids, and simply treat them as points. There are two main reasons behind this choice:

\begin{enumerate}
    \item 30 meters is a significantly higher resolution than we have for both the RWI map and the Voronoi partition. Also, as we mentioned above, the population cells are populated at most by 73 people because of the very high resolution. As a consequence, considering the areas of the population cells would add or remove at most a few dozen people, which would not make any significant difference in the computation of the aggregate RWI for the Voronoi cells.
    \item Treating points instead of polygons drastically decreases the computational cost. In our opinion, this additional computational cost does not add any significant value to the final outcome of the spatial aggregation.
\end{enumerate}

\begin{figure}[ht!]
    \centering
  \includegraphics[width=1\linewidth]{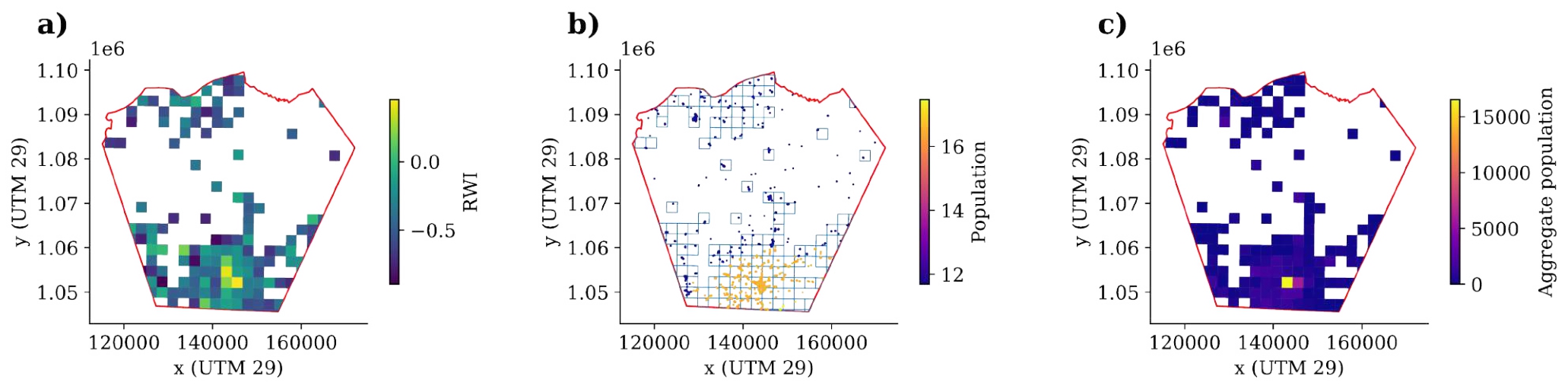}
\caption{\textbf{Spatial mapping.} (a) The spatial intersection between RWI cells (color-scaled) and a given Voronoi cell (red). (b) the population points inside each intersection. (c) The aggregate population inside each intersection. All the data shown here are either polygons or points georeferenced in UTM 29.}
  \label{figs9}
\end{figure}

Finally, we have the three datasets projected in the same coordinate system (UTM 29): the Voronoi polygons, the RWI square polygons, and the population points. To aggregate the RWI values over a given Voronoi cell $v$, we do the following (example results for a single Voronoi cell are shown in Fig. \ref{figs9}):

\begin{enumerate}
    \item We take the spatial intersections of the RWI cells that at least partially overlap with the Voronoi cell (Fig. \ref{figs9} (a)), and refer to them as the spatial intersections $A_v$ and their respective RWI value $R(A_v)$.
    \item We take the population points $p$ that lie within each RWI intersection $A_v$ (Fig. \ref{figs9}(b)).
    \item We take the sum of the population values $w_p$ inside each RWI intersection $A_v$ and consider it as the weight for the RWI aggregation, more precisely we define the weight $W(A_v)$ as:

    $$ W(A_v) = \sum_{p \; within \; A_v} w_p $$

    The population weights obtained for the example cell are shown in Fig. \ref{figs9}(c).

    \item To obtain one single aggregate RWI value $R_v$ for the Voronoi cell $v$, we average the RWI of the spatial intersections $R(A_v)$, weighting the average with the previously obtained population weights $W(A_v)$. More precisely:

    $$ R_v = \frac{\sum_{A_v} W(A_v) R(A_v) }{\sum_{A_v} W(A_v)} $$
    
\end{enumerate}

The final aggregated RWI map is shown in Fig. \ref{figs8}(c).

\section*{Additional information about home and SES inference}

As mentioned in the main text, we define a user's home location as the tower where most of the user's activity has been recorded during nighttime (9 PM-6 AM), lockdown, and curfew periods. We assume that users spend most of their time at home during these time windows and exploit this fact in the home location inference process: in the identification of the most frequent location, we double the representativeness of events recorded during lockdown or curfew. As mentioned in the main text, in all our analysis on segregation we do not consider home-to-home trips in the mobility network $G_M(t)$, as they would be clearly overrepresented during lockdown and curfew periods due to our home location inference methodology, thus they would lead to biases. For the call network analysis to be comparable, we also exclude calls/SMSs between users with the same home location in the social communication network $G_S(t)$. For the effects of these filtering see section "Effects of local spatial correlations" further in the SI Appendix.

We filter out all users with not enough statistics to infer their home locations (less than 2 geolocalized communication events) and with a spatial uncertainty ~\cite{vanhoof2018detecting} associated with their home location higher than 25 km. Lastly, we remove the most active $0.5\%$ of users in terms of incoming (respectively, outgoing) calls/SMSs, who at the same time have no outgoing (respectively incoming) communications, as they possess clearly anomalous communication behavior.  With this procedure and filtering, we infer the SES of 505,676 users.

We group users into 9 classes such that the number of users belonging to each class is as balanced as possible. Due to the arbitrary number of people with the same RWI value, classes cannot be perfectly balanced without assigning people with the same RWI label to different neighboring classes. Instead, we sort users by their RWI labels and segment them into classes with RWI boundaries chosen in such a way that the resulting classes are as balanced as possible in size and, at the same time, people with the same RWI are not assigned to different classes. The class division can be seen in Fig. 3F in the main text.

\section*{Population count and representativeness}

To assess if the spatial distribution of mobile phone users' home locations is representative of the actual population distribution, we compare it with aggregations of Facebook high-resolution density maps ~\cite{dataforgood} at various spatial resolutions. First, we make the comparison at the most fine-grained resolution, namely at the cell tower level. For mobile phone data, we simply count the number of users with inferred home locations at each tower. Regarding Facebook population count, we discard the spatial areas of data points (similarly to what we do in the spatial mapping pipeline), treat the data as points, and sum the population values of all points falling within the Voronoi cell of each tower. We then divide both population counts by the area of the cells to obtain population density values. The result is shown in Fig. \ref{figs21}(a) and (b). From Fig. \ref{figs21}(c) we can see that the two population densities are highly correlated ($\rho=0.80$) and that the mobile phone user home location population density is highly representative of the actual population density.

A further validation of the representativeness of our home location inference comes from the aggregation at lower spatial resolutions. In particular, we consider three census administrative levels, namely chiefdoms, districts, and provinces, that have increasingly lower resolution. Again, to compare mobile phone users and Facebook population we count the number of people within a given census cell. We can see from Fig. \ref{figs21}(d), (e), and (f) that our home location inference is highly representative ( $\rho = 0.94$ ) of the actual population at the most fine-grained census administrative level (chiefdoms), which has approximately half the resolution of cell towers (there are 405 towers and 207 chiefdoms). Finally, we get perfect correlations at both the district (Fig. \ref{figs21}(g), (h) and (i)) and the province level (Fig. \ref{figs21}(j), (k) and (l)).

We can conclude that our home location inference is highly representative of the actual population density distribution at the tower level and all administrative levels. A further validation of the representativeness of our sample, as mentioned in the main text, is the Gini coefficient of 0.38 that we obtain from the inferred RWI distribution of mobile phone users, which is comparable to the World Bank value of 0.36~\cite{worldbankgini}.

\section*{COVID-19 timeline and restrictions in Sierra Leone}

The first case of COVID-19 in Sierra Leone was confirmed on March 31, 2020. On April 1, after the second case was confirmed, the government announced a 3-d lockdown to be put in place on April 5-7. On April 9, the government announced a restriction on all inter-district travel for 14 days and a curfew from 9 PM to 6 AM. Also, face masks were strongly encouraged, only shops selling essential items were left open and people were asked to stay at home unless they had extremely urgent reason not to. There were also other types of restrictions, like school and workplace closures. The precise timeline and strength of the nine types of restrictions collected by ~\cite{hale2021global} are shown in Fig. \ref{figs18}, and the category codes are fully listed in Supp. Tables \ref{tabs2}-\ref{tabs10}

\section*{Reference periods and confounding events}

In our study, we analyze the effect of restriction policies by comparing social behavior during lockdown and curfew to two reference periods:

\begin{itemize}
    \item R1: March 22-24 2020
    \item R2: March 29-31 2020
\end{itemize}

We use these specific time windows because they are of the same length as the lockdown (the event of our greatest interest) and during the same days of the week (from Sunday to Tuesday). 

Despite being relatively close in time to the lockdown, they are both (especially R1) representative reference periods, in that they are still periods in which people were not significantly affected yet by the outbreak of the pandemic and by response policies. Indeed,  we can see from Fig. \ref{figs18} that the most socially impactful restriction policies (stay-at-home requirements, restriction to internal movements, restriction on gatherings) were put in place only after R2. However, relevant restrictions like school closures and workplace closures were already put in place during R2. The only restrictions that were put in place during R1 were the closure of public transport and the cancellation of public events, but it is unlikely that such measures had a strong effect on people's behavior. 

To assess if mobile phone behavior was different before our reference periods we can analyze the results obtained by Ndubuisi-Obi et al. ~\cite{ndubuisi2021using}. The authors also had access to CDRs data from Sierra Leone during the first wave of COVID-19 and used it to study compliance with mobility restrictions. However, their data spans a longer time window (February 2020 to May 2020). We can see from Fig. 3.3 in \cite{ndubuisi2021using} that mobility patterns at the national level did not change significantly in the months before the lockdown, and actually until the very beginning of the lockdown. This demonstrates that our reference periods are valid since a separate analysis of the same type of data shows that the temporal patterns were not significantly different before our reference periods. Moreover, they also show a very sharp decrease in mobility during the lockdown, similar to what we show in Fig. \ref{figs1} (see Section "Effects of lockdown on dynamics of social and mobility activities"). Their results add validity to our observations. 

Finally, we check if there were notable conflict events that occurred in that period and that could have affected the segregation patterns by analyzing the publicly available Armed Conflict and Location Events Database (ACLED) ~\cite{raleigh2010introducing}. In particular, we look at whether there were spikes of violence or protests in the same period of our observation. Fig. \ref{figs19} shows this data. We do not observe any particular increase in any type of events before, and during our observation period, so we can exclude that other events could have affected segregation. Also, looking specifically at our observation period (see Fig. \ref{figs20}), we can see that there are only two reported events with zero fatalities during the lockdown, so we can also exclude that such events have been relevant for segregation.

The only festivity during our observation period is the Eastern Weekend, from April 10 (Good Friday) to April 13 (Eastern Monday). This festivity falls during our CF time window and might have had an impact on segregation that we can not disentangle from our observations. However, we think that the restriction policies put in place during the curfew played a major role in driving segregation dynamics. Also, the possible confounding effect of Eastern during the curfew is of relative interest to us as our main focus in this study is the sharp and short-term impact on segregation observed during the lockdown period.

From this analysis, we can conclude that we can consider R1 and R2 as valid reference periods (especially R1) and that we do not notice other relevant events that could have affected significantly segregation during our observation period and especially the shock that we observed during the lockdown.

\section*{Rolling time window}

As explained in the caption of Fig. 1 in the main text for all calculations we use a 3-d rolling time window with 1-d shift. To make an example, if $t =$ April 10, $G_M(t)$ or $G_S(t)$ are the aggregate networks obtained from all the movements or communication events (resp.) recorded between April 9 and April 11. We make this choice to smooth the time series and to reflect the 3-d nature of the lockdown implemented by the Government of Sierra Leone during April 5-7. In this way, there is one point in the time series (April 6th) that incorporates all and only the interactions recorded during the lockdown.

However, the rolling time window can hide or smooth weekly patterns. We show the "raw" time series (without a rolling time window) in Fig. \ref{figs10}: in this case, every point refers to the segregation index computed from all the events recorded within a given day. We can clearly see that our main finding is not an effect of the rolling time window. Indeed, during the lockdown days, both curves show the same trends that we find in Fig. 1 in the main text, with segregation increasing significantly in the mobility network and decreasing significantly in the communication network, with respect to reference periods. However, we can also clearly see that the rolling time window is hiding some weekly patterns related to work and school routines (working days in Sierra Leone are from Monday to Friday). From the mobility curve in Fig. \ref{figs10} we can see that during reference times mobility segregation is higher during weekdays than during weekends. The same can be said, to a lesser extent, for communication. We can link this observation to the fact that people during the weekend have normally more free time to explore different places and to communicate with a different set of people than during work days. Also, we can distinguish Sunday during the lockdown in the communication curve much more than in reference periods, while we can not in the mobility curve. This means that while the weekly pattern of social mixing is flattened in the physical space by stay-at-home requirements, it is amplified in the communication space.

\section*{Effects of lockdown on dynamics of social and mobility activities}

To identify the effects of lockdown on the mobility and social behavior of different SE groups, we compute the daily evolution of mobility and social activities. In each group, we consider the average travel distance (among all travels recorded at time $t$) for mobility and the average number of communication events per person (among all active users at time $t$) for communication. These measures (in Fig. \ref{figs1}(a),(b)) show clear patterns of behavioral change in response to the lockdown (LD). In general, they show significantly reduced travel distances and communication activities for everyone during this period but they also signal that people from different SE classes have evidently different capacities to adjust to the emergency. In terms of travel distance, during the pre-intervention period (weeks of R1 and R2 in Fig. \ref{figs1}(a),(b)), the travel patterns of higher SE classes followed regular weekly cycles while they traveled the shortest distances (on average $\sim 14 km$) on a daily basis. Since people of higher SES are likely white-collar workers and occupy office jobs, they could adjust effectively to the lockdown and reduce their daily travel distance close to the minimum. On the other hand, people from lower SE classes had similar travel patterns on each day of the week and traveled significantly longer distances ($\sim 37 km$) during the reference period. During the lockdown, they only reduced their travel distance to $\sim 20km$. Although this is a larger relative reduction, it is far from the level that was possible for people from higher SE classes. During the curfew period, we see a steady return to close to normal travel distances, although none of the classes return fully to their pre-lockdown travel behavior. Notice that the large difference observed between the average travel distance of poor and rich people is also to be ascribed to the difference in size between the Voronoi cells of rich and poor areas. % but evidently richer people could keep to adjust better to continuing emergency.

Similar SE patterns characterize the overall communication dynamics of SE classes, as shown in Fig. \ref{figs1}(b). People from lower SE classes perform fewer number of communication events ($\sim 4$) on reference days and depict no weekday-weekend patterns in their dynamics. Nevertheless, during the lockdown, they reduce their communication activities the least, by $\sim 25\%$, while making the minimum number of calls in the population. In contrast, people from higher SE classes have an overall higher number of communications, that show clear weekly cycles with smaller activity during weekends. During lockdown, their communication activities fall relatively the most ($\sim 40\%$), yet they call the most during this period. This suggests that richer people could adjust more in terms of mobile phone communication volume, which in turn likely impacts the structure of their personal network.

Consequently, the lockdown interrupted both the social and travel dynamics of people but impacted people from various SE classes differently. While the reduction in mobility was expected due to the stay-at-home order ~\cite{engle2020staying,schlosser2020COVID}, similar changes in social communication are surprising as these mobile communications were not limited by similar constraints.

\section*{Individual segregation}

We computed the individual segregation index for each nodes in each network for each observation period (R1, R2, LD, and CF). Subsequently, we measured the $P(r_u(t))$ distribution of this index separately for each SE class and computed the median and standard deviation of this distribution for each period and SE class. This is summarized in Supp. Table \ref{tabs1} for the $G_S(t)$ and $G_M(t)$ networks separately. For the interpretation of this table see the main text.

\section*{Segregation measured with entropy}

Segregation is often measured in terms of diversity, through entropy-like metrics ~\cite{eagle2010network,moro2021mobility,yabe2023behavioral}. The principle behind these metrics is that the more entropic (i.e. diverse) the SES distribution of a given person's contacts, the less segregated the person is. In terms of mobility, the more entropic the SES distribution of places visited by a given person (or of people visiting the same places as a given person), the less segregated the person is. Despite being widely used metrics, we choose not to work with entropy-like metrics and to use assortativity for two main reasons:

\begin{itemize}
    \item The homophily phenomenon: while entropy only considers the socioeconomic status (SES) distribution of an ego’s neighbors without referring it to the ego’s SES, assortativity explicitly measures the correlation between an ego’s SES and its neighbors’ SES. To make an example, if a node that belongs to the poorest class has connections only with the poorest class or only with the richest class, the entropy will be the same (zero). On the other hand, assortativity gives us two opposite outcomes (in this case, segregation and anti-segregation). We believe it is better to use assortativity in our case because it is more nuanced, and captures the homophily mechanisms behind segregation.
    \item Data sparsity: given the typical long-tail distribution of user’s activity (see Fig. \ref{figs11}), for most individuals, we only observe a few links in a single time window. As such, most people will be assigned low entropy values, if not exactly 0 (see Fig. \ref{figs13}). Global assortativity is not affected by this issue because it measures one single correlation coefficient between all nodes’ SES and their neighbors’ SES. Moreover, individual-level segregation overcomes this issue, since it captures the correlation between an ego’s network and its local network, by assigning exponentially decreasing weights to distant nodes.

\end{itemize}

One advantage to working with entropy in mobility, however, is that we can compute individual values not only at the location level (the diversity of users visiting a location) but also at the user level (the diversity of places visited by a user). 

In this section, we replicate the segregation analysis with an entropy-like measure of diversity and analyze explicitly the two problems mentioned above. Likewise the individual-level analysis, we assign each user and each location to one out of nine socioeconomic classes. Given the set of places visited by a user $u$, we compute the empirical probability that the user visits places of a given socioeconomic class $P_m(u; c)$, with $c \in \{1,...,9\}$, by normalizing the frequency of visits to places of each socioeconomic class. We then define the user mobility diversity $D_m(u)$ as:

$$ 
D_m(u) = - \sum_{c=1}^{9} P_m(u; c) \log P_m(u; c) 
$$

Similarly, given the set of users visiting a place $p$, we compute the empirical probability that the place is visited by users from a given socioeconomic class $P_m(p; c)$, with $c \in \{1,...,9\}$, by normalizing the frequency of visits from users of each socioeconomic class. We then define the place mobility diversity $D_m(p)$ as:

$$ 
D_m(p) = - \sum_{c=1}^{9} P_m(p; c) \log P_m(p; c) 
$$

Finally, given the set of users with whom a user $u$ communicates, we compute the empirical probability that the user communicates with users from a given socioeconomic class $P_s(u; c)$, with $c \in \{1,...,9\}$, by normalizing the frequency of communication with users from each socioeconomic class. We then define the user social communication diversity $D_s(u)$ as:

$$ 
D_s(u) = - \sum_{c=1}^{9} P_s(u; c) \log P_s(u; c) 
$$

Since we have 9 socioeconomic classes, the three metrics are all bounded between $0$ (connections to a single socioeconomic class) and $\log 9 \sim 2.20$ (equal frequency of connections with all socioeconomic classes). The temporal evolution of the mean values of $D_m(u)$, $D_m(p)$ and $D_s(u)$ is shown in Fig. \ref{figs12}. 

Regarding mobility, we can see that places' diversity is overall significantly higher than users' diversity. The main reason behind this difference is data sparsity since single users have much fewer connections than single locations (the same number of user-to-location links is shared among 505,676 nodes on the user side and 405 nodes on the location side). Therefore, for many users $D_m(u) = 0$ simply because they have very few recorded events. We can also see that both diversities strongly decrease during lockdown, which is the analogous observation we do with assortativity in Fig. 1 in the main text (decreasing diversity is equivalent to increasing assortativity in terms of segregation).

Regarding social communication, the diversity measure allows us to do a direct comparison with mobility at the user level. We can see that on average the set of of contacts of a user is more diverse than the set of visited places. This is likely due to the constraints of physical distance, which are clearly stronger in mobility than in communication (it is easier to communicate with someone living far away than to visit a faraway location, and at the same time faraway places are more diverse in terms of RWI). During lockdown, in Fig. 1 in the main text, we observe a decreasing assortativity, which would correspond to an increasing diversity. However, we can see that diversity slightly decreases during lockdown. The reasons behind this discrepancy are the two issues with entropy-like metrics that we mentioned initially, namely data sparsity and the inability to capture homophily. To demonstrate it, for each time step, we analyze only the set of users with $D_s(u)=0$, which are always the majority due to data sparsity (see Fig. \ref{figs13}). At a given time step, these users communicate exclusively with a single socioeconomic class. However, entropy alone is not able to tell us if this single socioeconomic class is similar to the user's socioeconomic class and if this changes with time. We can measure this effect if we group users with $D_s(u)=0$ by their socioeconomic class and look at the distribution of their aggregate interactions. For example, we take all users with $D_s(u)=0$ at a given time belonging to class 1 (the poorest). We look at the socioeconomic class distribution of their aggregate interactions (which indicates how many of these users interacted exclusively with class 1, with class 2, ...). Temporal changes of this distribution indicate some homophily change that entropy can not capture because users are anyway assigned with $D_s(u)=0$. The two distributions for R1 and LD are shown in Fig. \ref{figs15}(a). We can see from the figure that during LD more users are communicating exclusively with richer (i.e. more distant) socioeconomic classes than in R1. To measure this effect we calculate the entropy of these distributions for every day. From Fig. \ref{figs15}(b) we can see that during LD the entropy is higher than during reference periods, and this observation holds for every class (see the right panels in Fig. \ref{figs15}). This indicates that in every socioeconomic class during LD, more users are communicating exclusively (i.e. with $D_s(u)=0$) with more distant socioeconomic classes than during reference periods. This means in turn that there is a decrease in homophily during lockdown that entropy-like metrics like diversity are not capturing, and given the predominant number of users with $D_s(u)=0$ (see Fig. \ref{figs13}) this difference is determinant of the discrepancy that we observe between assortativity and diversity in the social communication network. 

Regarding mobility, the same analysis can be applied to users with $D_m(u)=0$ and it is shown in Fig. \ref{figs14}. However, in this case, the effect goes in the same direction as the mean diversity, which means that in mobility there is an even stronger increase in homophily and segregation than what is captured by entropy-like metrics. 

\section*{Effects of local spatial correlations}

As discussed in the main text, in the analysis of segregation dynamics we removed all home-to-home events from both $G_M(t)$ and $G_S(t)$ networks. Here we check that the main results of the segregation analysis are not altered by these removals. In Fig. \ref{figs2} we show the analogous panels of Fig. 1 in the main text, without removing home-to-home events. We can see that the presence of home-to-home events induces the sharply diagonal shape of the distributions in panel (c) of Fig. \ref{figs2}, with a network assortativity coefficient $\rho$ close to 1, visible in panel (e). However, we can see that the main result claimed in the main text (the opposite direction of the network assortativity change of $G_M(t)$ and $G_S(t)$ during the lockdown) is clearly visible also in Fig. \ref{figs2}, and hence not determined by the presence/absence of home-to-home events.

Also, in Fig. \ref{figs3} we show the analogous of the bottom panels (D-F) of Fig. 2 in the main text for the social communication network $G_S(t)$, without the removal of communication events between people with the same inferred home location. For this result, we can not reproduce the results in the main text for the mobility network $G_M(t)$ without the removal of home-to-home travels as the local assortativity index $r_u(t)$ is defined only for a network with no self-loops. Since nodes are locations in the mobility network $G_M(t)$, home-to-home links are trips that start and end in the same node (hence self-loops). On the other hand, in the social communication network $G_S(t)$ home-to-home links are communication events between different nodes that live in the same place (hence are not self-loops). We can see from panels (a), (b), and (c) in Fig. \ref{figs3} that the results change only slightly from the bottom panels of Fig. 2 in the main text. The main message remains the same, with a clear dependence on the SES of the relative shift of the distribution $P(r_u(t))$ during intervention periods (panel c). The majority of classes become less segregated than during reference periods (the poorer the class, the higher the segregation decrease) and only the richer classes become more segregated, leaving the global segregation to decrease just like in Fig. 3E in the main text.

\section*{Effects of professional activities}

As we discussed in the main text, one of the possible confounding effects that can induce the reduction of mobility and social communication activities during the lockdown period is the lack of professional communications due to interrupted businesses and closed offices. Here we investigate this factor by separating the mobility and call activities in our data for office hours, between 9 AM and 7 PM, and out-of-office hours (7 PM-9 AM) and recompute the daily segregation indices for the networks constructed from events of mobility and communication falling within these periods, with the same resolution we used in the main text (three-day time window with daily shift).

We can see from Fig. \ref{figs4} that the effects of professional activities are not the driver of the segregation changes during the lockdown, neither for the mobility nor for the social communication network. Indeed, separating the activity during office hours from the one during out-of-office hours, we do not find significant differences in the segregation dynamics. The only relevant difference is between the two curves of social communication segregation, where segregation during working hours is systematically higher than during the remaining times. Nevertheless, a very similar segregation decrease appears during the lockdown in both curves, proofing that these reorganization patterns were not due to the interruption of professional communications.

\section*{Effects of spatial and network correlations}

In the main text, we have shown signs of SE segregation both in mobility and in the social network. However, also simple random network models might produce positive segregation values, thus making our observations ambiguous from the point of status homophily. Indeed, segregated configurations might result from the convolution of multiple factors, among which status homophily provides only one explanation. To assess the significance of status homophily in the assortative network formation, we identify three main confounding factors and measure separately their contributions to the observed segregation levels.

\subsubsection*{Physical distance} It is known that distance has a strong determining effect on spatial network formation~\cite{barthelemy2011spatial}. On one hand, nearby places might host populations with similar SE profiles. On the other hand, nearby places are likely to have more mobility or even communication connections among each other. The convolution of these two effects could explain the observed segregation patterns. To measure the impact of physical distance, we consider a gravity model~\cite{barthelemy2011spatial}, where the number of connections $W_{ij}$ between places $i$ and $j$ is only determined by the number of people living in the two places ($N_i$ and $N_j$) and the distance $d_{ij}$ between them, according to the law:

\begin{equation}
W_{ij} = C \frac{N_i^{\alpha} N_j^{\beta}}{d_{ij}^{\gamma}}
\label{eq1}
\end{equation}

The three exponents $\alpha$, $\beta$, and $\gamma$, and the constant $C$ are fitted at every time $t$ from the data with an ordinary least square (OLS) regression. The values of the exponents can be seen in Fig. \ref{figs6}. As evident from Fig. \ref{figs5}, assortative indices computed from the gravity model structures are relatively high but do not fully reproduce the level of segregation observed in the real network. Consequently, distance effects contribute to the emergence of the observed segregation but they do not fully explain them.

However, we can see that gravity model curves have similar temporal tendencies as the empirical ones, which is due to overfitting. Indeed, we fit a gravity model per day, instead of fitting a model for the full time-period. We say the model is overfit because we are fitting the three exponents of the gravity model at every timestep (i.e. every day), and predicting the weights of the same set of links on which the model was trained. The three exponents are supposed to be universal, but as seen in Fig. \ref{figs6} they fluctuate over time and significantly change during the lockdown. The purpose of this analysis is to show that even the most overfitted gravity-like model, despite having strong explanatory power, is not able to reproduce the same level of segregation, implying that there is also a social preference mechanism that comes into play.

\subsubsection*{SE network correlations} Next, we consider the impact of SE status and network correlations on the observed assortativity correlations to see whether they are contributing at all to the emergent network segregation patterns. To test the impact of the RWI distribution, we randomly swap the RWI labels among nodes (being people in $G_s(t)$ or places in $G_m(t)$), keeping the overall distributions and the network structures fixed ~\cite{gauvin2022randomized}. In our implementation we perform 100 random swap iterations and consider the mean assortativity value, for every day, resulting from such iterations. This procedure destroys the network-SES correlations and results in assortativity indices close to zero (see Fig. \ref{figs5}). Consequently, SE correlations are important in the emergence of network segregation patterns, as their removal leads to the vanishing of the observed patterns.

\subsubsection*{Network structure correlations} Finally, our last goal is to verify how much degree heterogeneities are important for the emergence of the segregation patterns in the networks. We test these effects by using configuration network models where we swap the ending nodes of a pair of edges to remove any structural correlations from the network ~\cite{milo2003uniform}. Note that this method keeps intact the degrees (number of connections) of nodes and the degree-SES label correlations. Also in this procedure we perform 100 random swap iterations and consider the mean assortativity value, for every day, resulting from such iterations. Results are shown in Fig. \ref{figs5}, where the assortativity indices measured in the configuration networks appear around zero, indicating that the removal of structural correlations completely destroys network segregation, thus the degree distribution and degree-label correlations do not contribute at all to the original observations neither in case of the mobility or the social communication network.

\section*{Individual gravity model and spatial aggregation}

Given the strong explanatory power of the gravity model that we observe in Fig. \ref{figs5}, we replicate here an analogous analysis for individual segregation. However, applying the same methodology to the individual analysis is not straight-forward and presents the following difficulties:

\begin{enumerate}
    \item Fig. 2 in the main text shows segregation patterns (assortativity) on the individual level. On this level, individual units in the social communication network are people. For a gravity model (which predicts connections between locations), individual units are places. For the global network assortativity this is not an issue, since we obtain the assortativity from a list of RWI tuples (caller’s and callee’s RWI), which do not change if we aggregate users at a location level (RWI of caller’s and callee’s home location). However, to compare the two we need to ensure that we compare them on the same scales. To construct a gravity-like model for social connections we would first need to aggregate users at the location level, predict links with the gravity model, and somehow map the links back to individual users. The last step is especially critical and would require some non-well-defined random link redistribution rules among users with the same home location. Results will then strongly depend on the type of rule that we decide and it would be hard to assess the role of the gravity model.
    \item Individual level assortativity is defined for unweighted networks, which means that in the data we consider all links, with at least one observed connection, equally. The gravity model allows us to predict a weight for any pair of places, only based on the population of the two places and the distance between them. This means that the predicted network is always a complete weighted network. Removing the weights would turn it into a simple complete (i.e. fully connected) network, making it impossible to see any change in time or space.
\end{enumerate}

Here we try to overcome these obstacles and to come up with a relatively simple individual gravity-like validation that can be applied to both the mobility and the communication networks. In particular, we overcome the first issue by avoiding the last step (the reverse mapping from places to users) and comparing social networks and gravity-like networks at the location level rather than the user level. More precisely, given a network $G_S(t)$ at any time $t$, we first aggregate the network at the location level by mapping every link $(u,v)$ to $(h(u), h(v))$, where $h(u)$ and $h(v)$ are the home locations of users $u$ and $v$, respectively. Secondly, we fit a gravity model \ref{eq1} from the location level link weights. Then, to overcome the second issue, we sort links according to the weight predicted by the gravity model and keep only the first $m$ links, where $m$ is the number of links observed in the empirical network aggregated at the location level. In this way, both networks (the empirical and the gravity-like) have the same number of links. Finally, we remove weights from both networks and compute the individual assortativity for all nodes in both networks. We do this for all time windows R1, R2, LD, and CF. Regarding the mobility networks, we follow the same procedure except for the first step which is not necessary since mobility networks are already defined at the location level.

The results are shown in Fig. \ref{figs16}. For every socioeconomic class, we show the mean value of the individual segregation distribution with the standard deviation, for the four time windows mentioned above and for both the empirical and the gravity-like networks, for mobility (Fig. \ref{figs16}(a)) and social communication at the location level (Fig. \ref{figs16}(b)). We can see that the mean empirical level of individual segregation is always higher than the one we measure with the gravity model, for both mobility and social communication, and for all periods and socioeconomic classes (in most cases by more than one standard deviation). However, the distance between the empirical values and the ones obtained from the gravity model is higher for mobility. We can also see that the segregation in the gravity-like networks does not show any significant temporal trend, such that we can exclude the role of spatial effects in this analysis. 

Compared with the distributions in Fig. 2D-E in the main text, we can see that individual segregation at the location level in the social communication network takes lower values than at the user level. Moreover, we notice that aggregating social networks at the location level hides the increase in segregation for the highest socioeconomic classes observed at the user level (as can be seen in Fig. 2F in the main text). Both these observations suggest that the spatial aggregation of the social communication network implies some loss of information. Looking deeper into this effect, we find that by applying some level of thresholding on the weak links (i.e. with a low number of connections), we are able to retrieve the observation also at the location level. By thresholding, we mean considering only links with a weight higher than a given threshold before removing weights from the network to calculate individual segregation. The results for different threshold values are shown in Fig. \ref{figs17}. We can see that for higher threshold values, the mean segregation in social communication decreases only for the lower socioeconomic classes during LD, while it increases for the highest ones. On the other hand, thresholding does not impact significantly the observations in the mobility networks, except for amplifying the differences between classes during LD, with the richest classes experiencing almost no change in segregation compared to R1 and R2 for high threshold values. We can conclude that while the observations in the mobility networks are very robust against weak connections, for the social communication networks we need to apply some link weight thresholding to balance the information loss that we necessarily experience by aggregating from individual to location level.

\newpage

\newpage

\newpage

\newpage

\begin{figure}[ht!]
    \centering
  \includegraphics[width=0.5\linewidth]{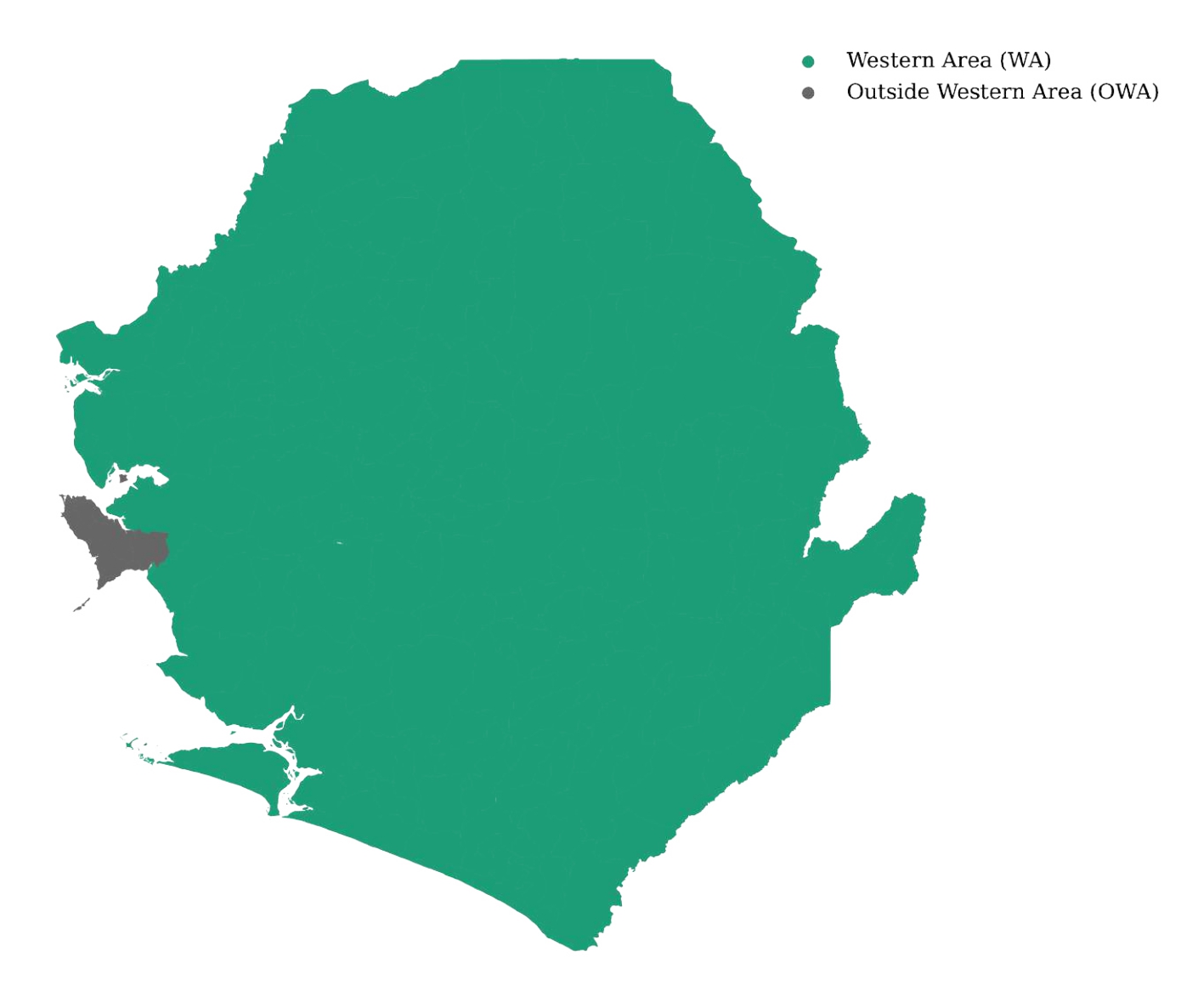}
\caption{\textbf{Spatial division.} Western Area (WA) in grey and Outside Western Area (OWA) in green.}
  \label{figs23}
\end{figure}

\newpage

\begin{figure}[ht!]
    \centering
  \includegraphics[width=0.9\linewidth]{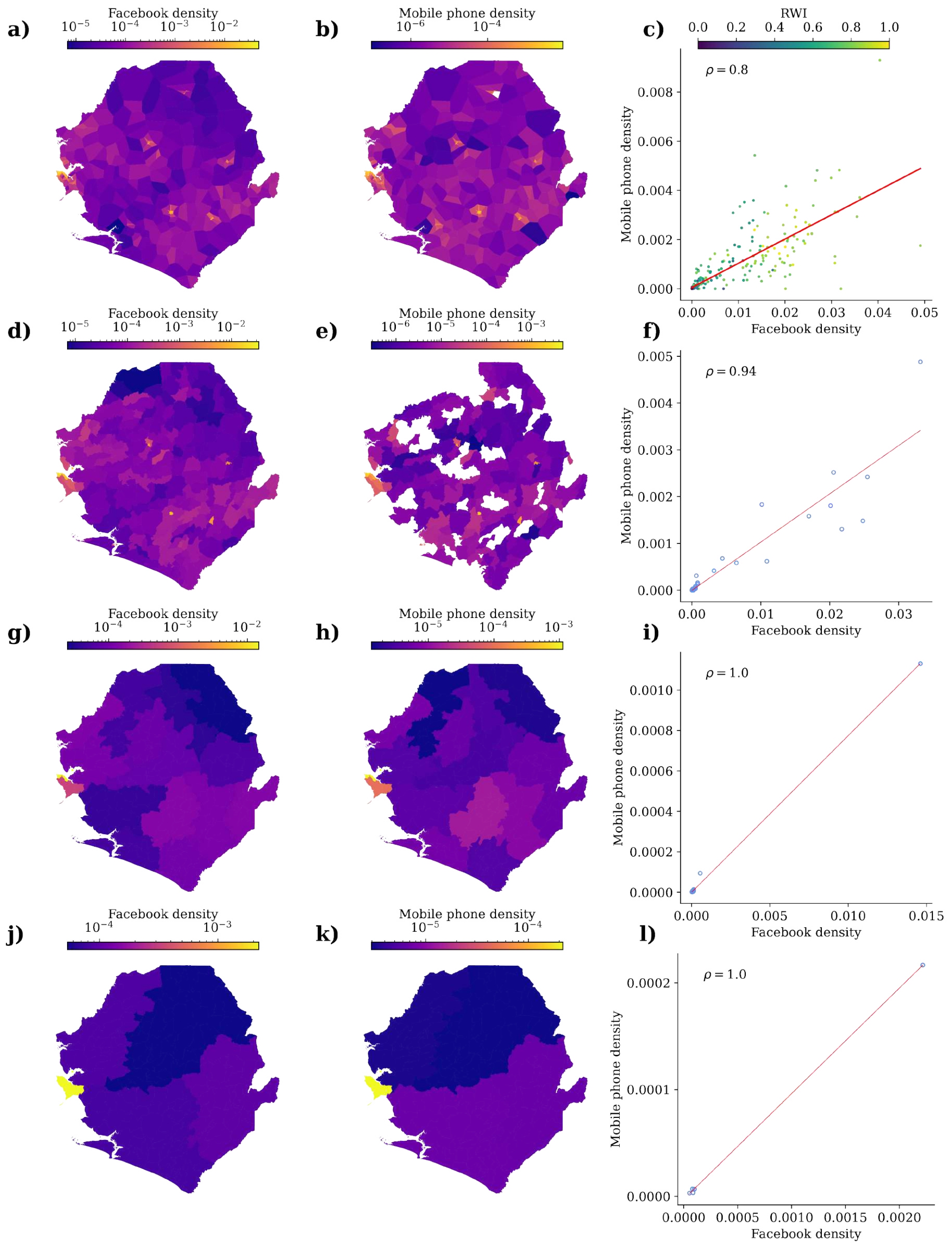}
\caption{\textbf{Population density validation.} (Left panels) Population density in each Voronoi cell (a), chiefdom (d), district (g), and province (j) from Facebook high-resolution population density data ~\cite{dataforgood}. (Middle panels) Population density in each Voronoi cell (b), chiefdom (e), district (h), and province (k) from mobile phone user's home locations. (Right panels) Relation between the population density from Facebook data (left panels) and mobile phone users (middle panels) at Voronoi cell (c), chiefdom (f), district (i), and province (l) level. The red lines are the results of OLS linear regressions. $\rho$ is the Pearson correlation coefficient.
}
  \label{figs21}
\end{figure}

\newpage

\begin{figure}[ht!]
    \centering
  \includegraphics[width=1\linewidth]{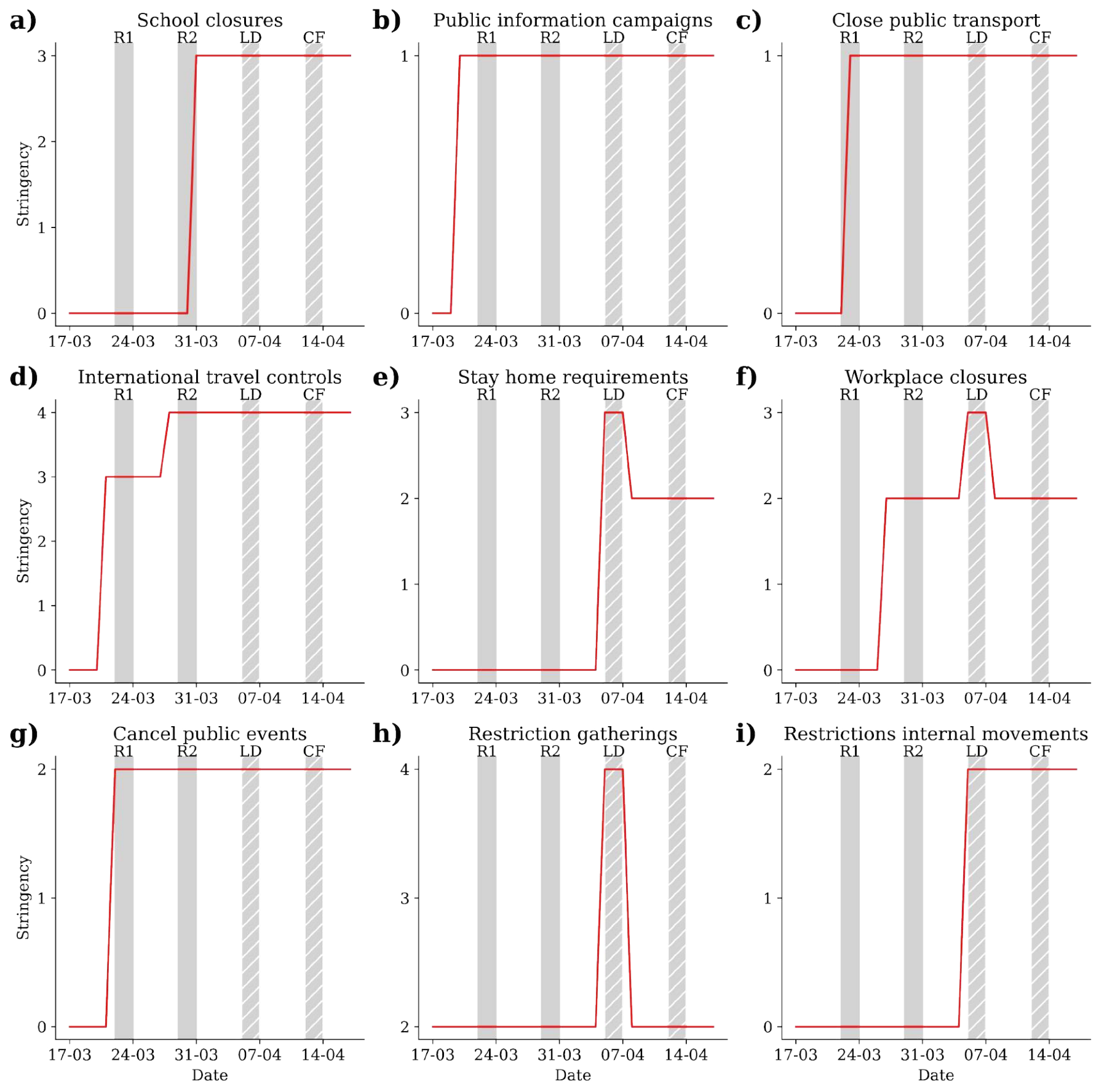}
\caption{\textbf{COVID-19 restriction policies in Sierra Leone.} Timeline of 9 restriction measures put in place in Sierra Leone in response to the first wave of COVID-19. The meaning of each category can be found in the Tables section or the original paper ~\cite{hale2021global}. 
}
  \label{figs18}
\end{figure}

\newpage

\begin{figure}[ht!]
    \centering
  \includegraphics[width=1\linewidth]{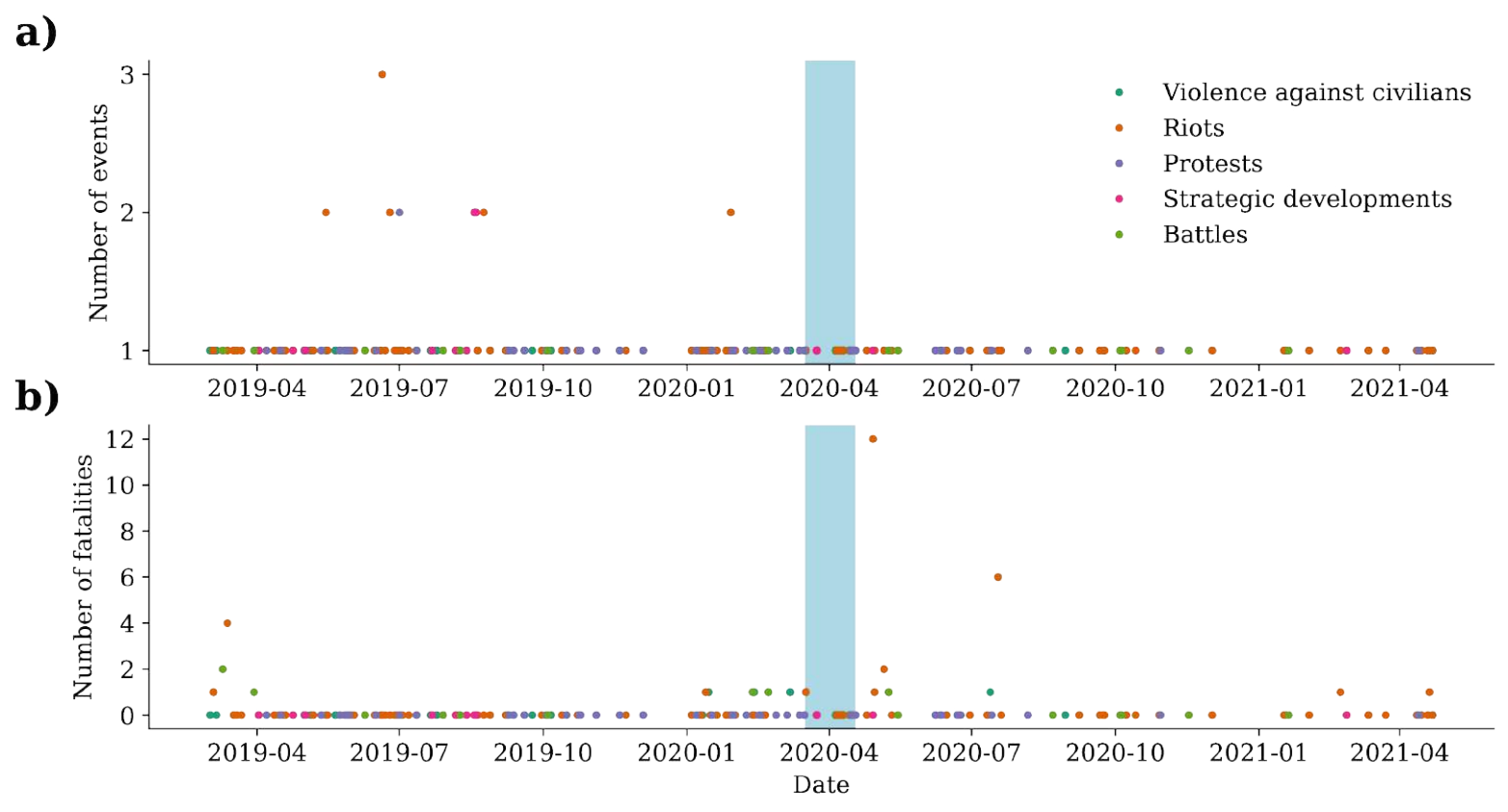}
\caption{\textbf{Armed Conflict Location and Event Data.} (a) Number of reported events from different categories. (b) Number of reported fatalities in the events from different categories. The observation period of our study (March 17 - April 17, 2020) is highlighted in light blue in both figures.
}
  \label{figs19}
\end{figure}

\newpage

\begin{figure}[ht!]
    \centering
  \includegraphics[width=1\linewidth]{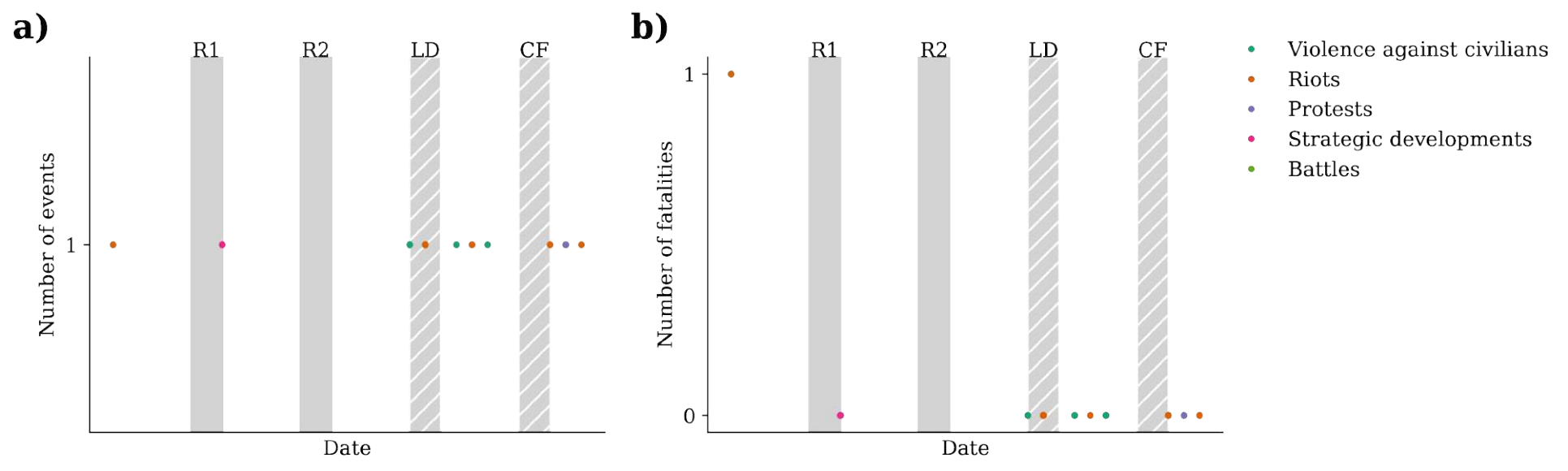}
\caption{\textbf{Armed Conflict Location and Event Data (restricted).} Same as Fig. \ref{figs19} but restricted to the observation period of our study. (a) Number of reported events from different categories. (b) Number of reported fatalities in the events from different categories. The time windows of interest (R1, R2, LD, CF) are highlighted.
}
  \label{figs20}
\end{figure}

\newpage

\begin{figure}[ht!]
    \centering
  \includegraphics[width=0.8\linewidth]{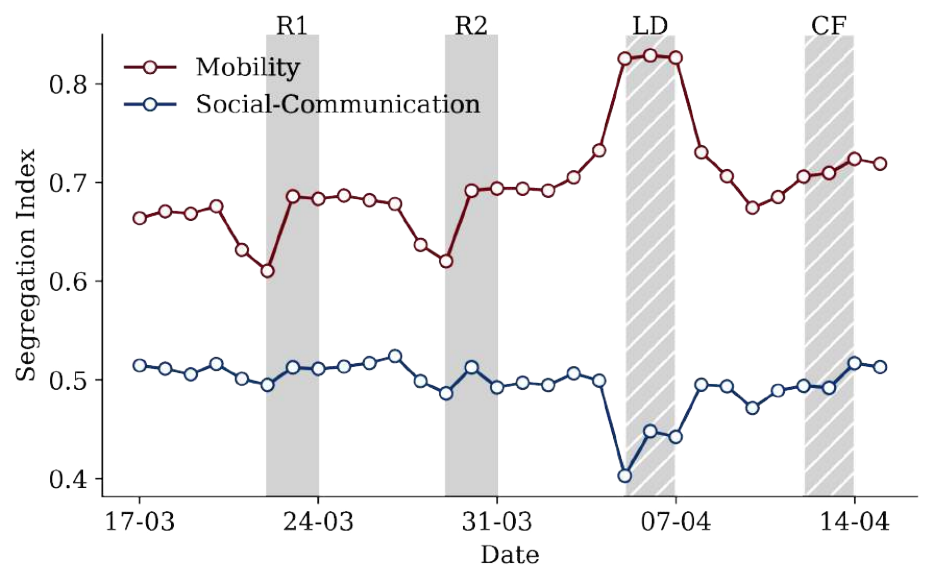}
\caption{\textbf{Raw segregation time series.} Segregation dynamics for the mobility (red) and social communication (blue) networks, with no rolling time window (every point refers to all events recorded within a given day). }
  \label{figs10}
\end{figure}

\newpage

%%% Each figure should be on its own page
\begin{figure}[ht!]
  \centering
  \includegraphics[width=\linewidth]{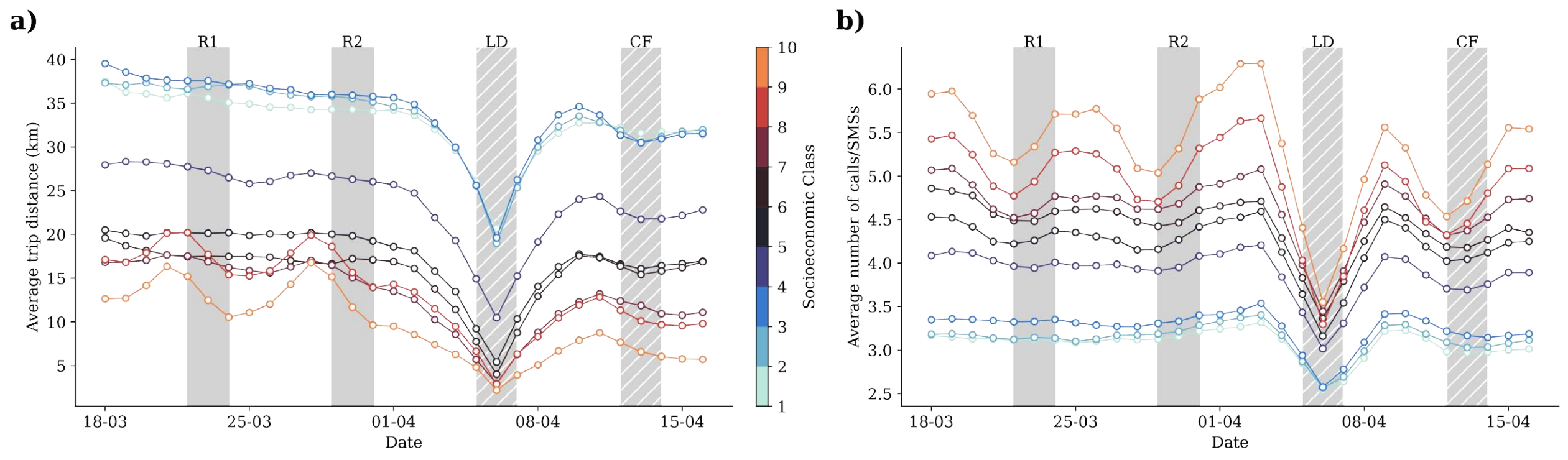}
  \caption{\textbf{Impacts of interventions on social and travel patterns.} Evolution of (a) average trip length and (b) number of mobile communication events for people belonging to different SE classes.}
  \label{figs1}
\end{figure}

\newpage

\begin{figure}[ht!]
  \centering
  \includegraphics[width=\linewidth]{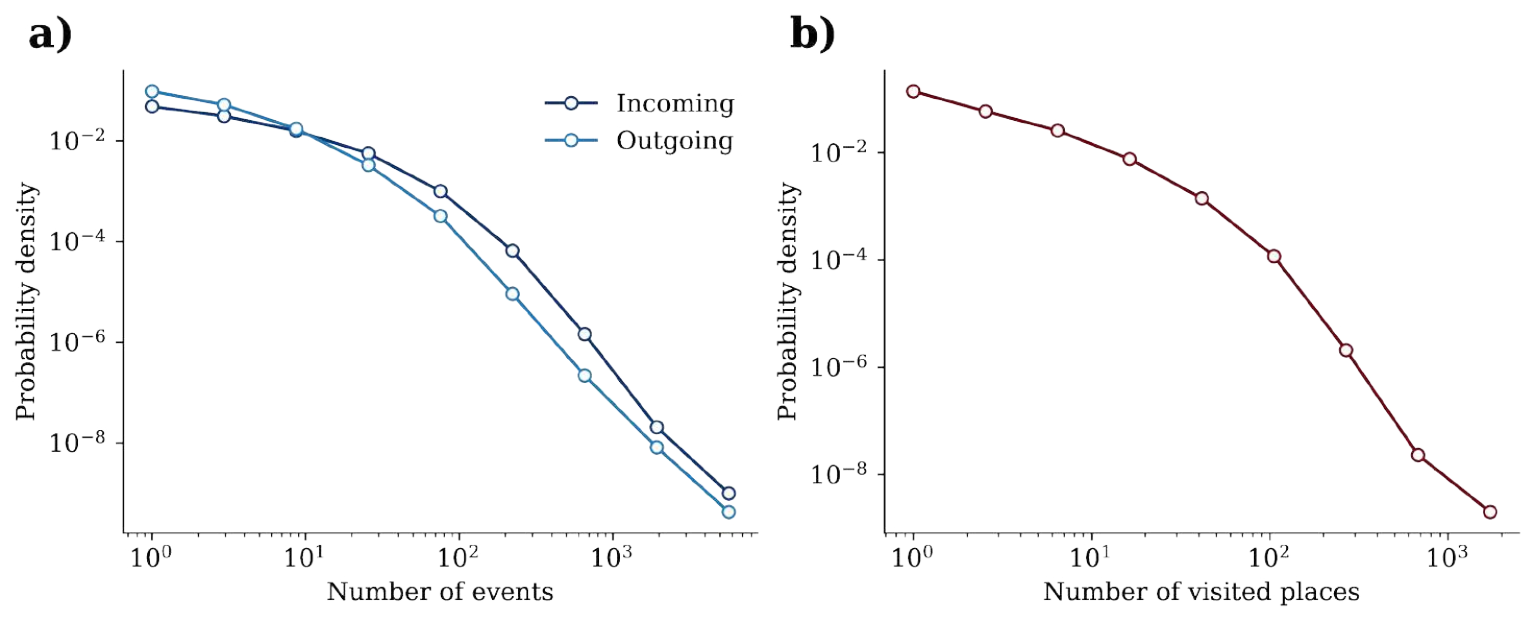}
  \caption{\textbf{Event distribution} (a) Distribution of the number of incoming and outgoing communication events per user. (b) Distribution of number of visited places per user.}
  \label{figs11}
\end{figure}

\newpage

\begin{figure}[ht!]
  \centering
  \includegraphics[width=0.8\linewidth]{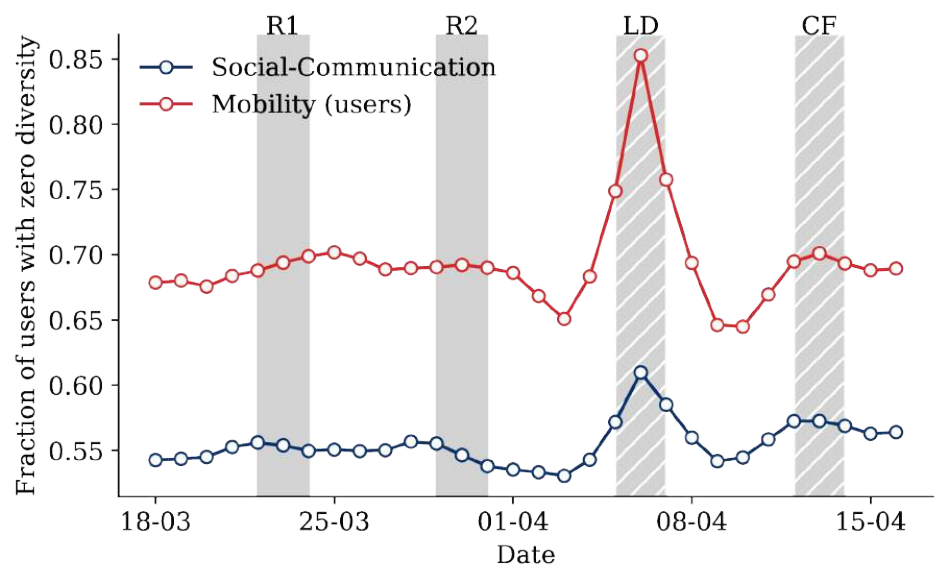}
  \caption{\textbf{Users with no diversity.} Fraction of active users with $D_m(u)=0$ (red) or $D_s(u)=0$ (blue).}
  \label{figs13}
\end{figure}

\newpage

\begin{figure}[ht!]
  \centering
  \includegraphics[width=0.8\linewidth]{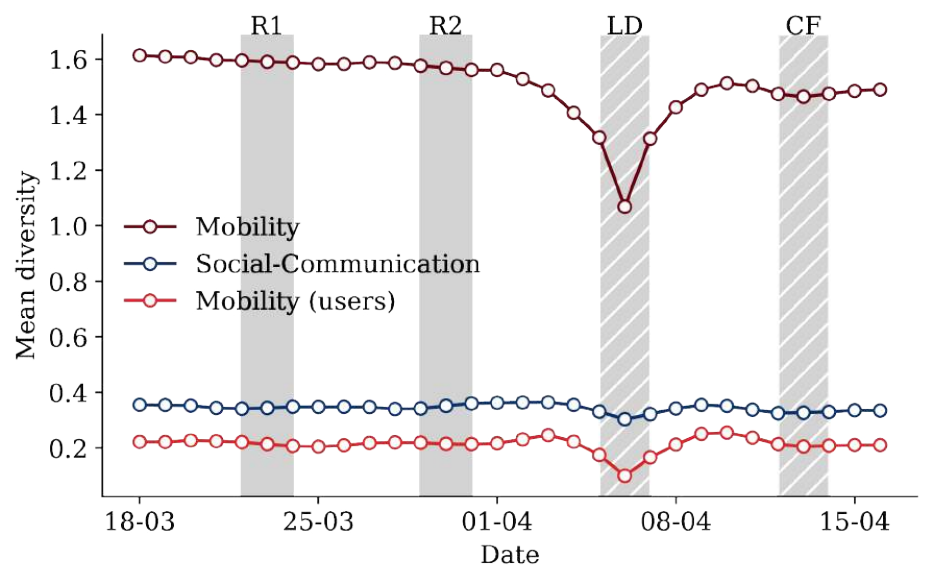}
  \caption{\textbf{Mean diversity.} Dynamics of mean diversity for mobility in terms of locations (dark red) and users (light red) and for social communication (blue).}
  \label{figs12}
\end{figure}

\newpage

\begin{figure}[ht!]
  \centering
  \includegraphics[width=0.6\linewidth]{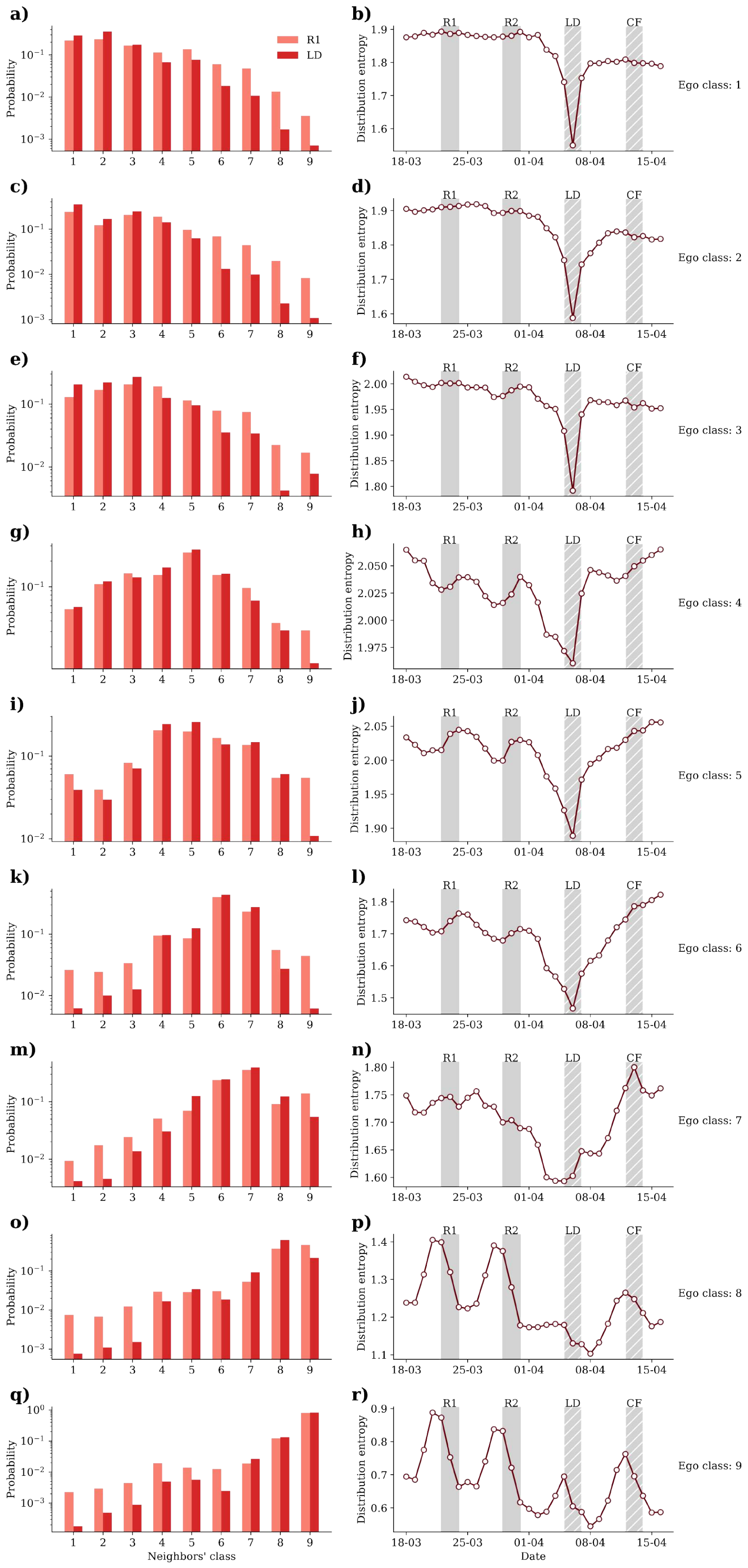}
  \caption{\textbf{Homophily changes not captured by diversity in the mobility network.} (Left figures) Comparisons between the socioeconomic distribution of the aggregate interactions of users with $D_m(u)=0$ during R1 and LD, for all classes from class 1 (a) to class 9 (q). (Right figures) Dynamics of the entropy of the socioeconomic distribution of the aggregate interactions of users with $D_m(u)=0$, for all classes from class 1 (b) to class 9 (r). }
  \label{figs14}
\end{figure}

\newpage

\begin{figure}[ht!]
  \centering
  \includegraphics[width=0.6\linewidth]{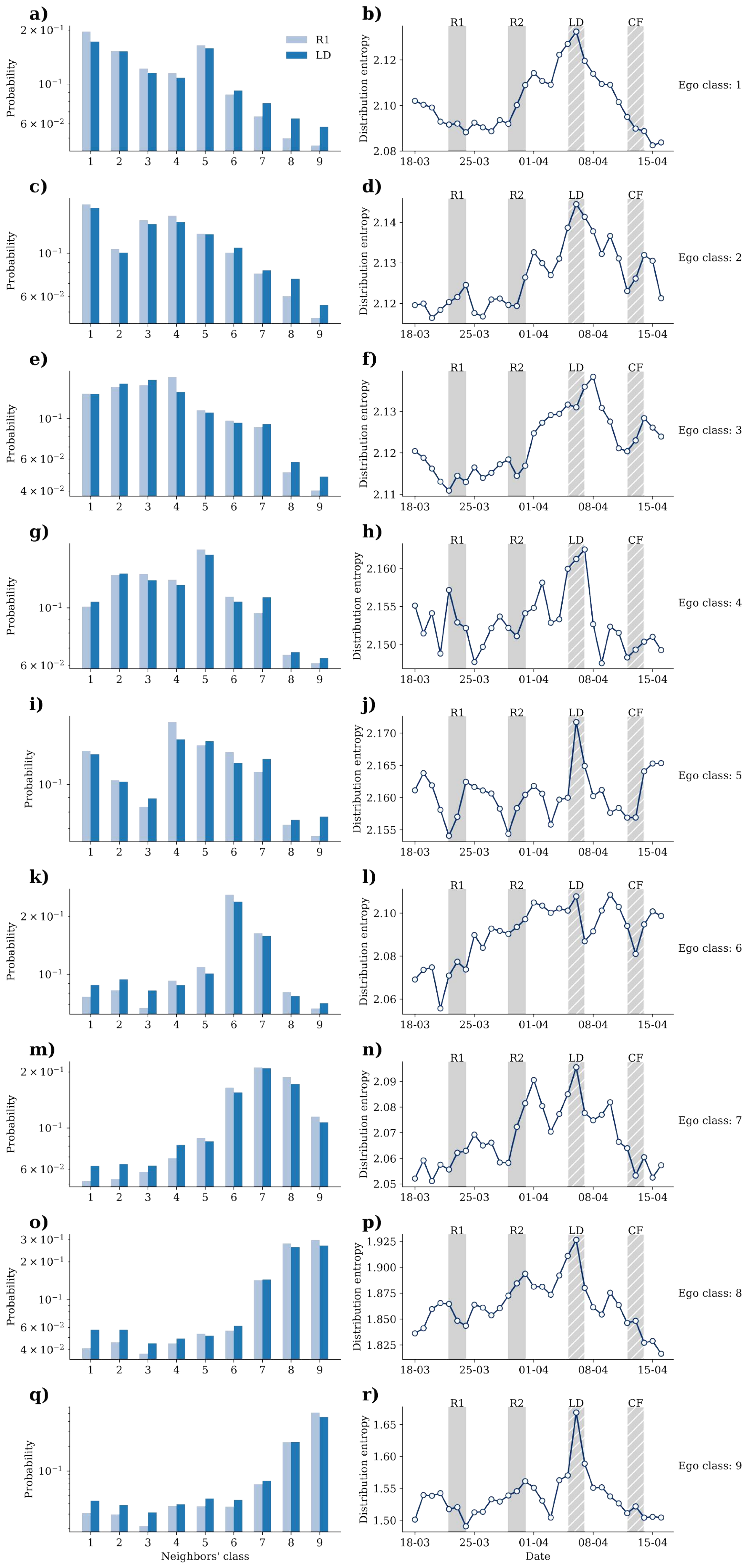}
  \caption{\textbf{Homophily changes not captured by diversity in the social communication network.} (Left figures) Comparisons between the socioeconomic distribution of the aggregate interactions of users with $D_s(u)=0$ during R1 and LD, for all classes from class 1 (a) to class 9 (q). (Right figures) Dynamics of the entropy of the socioeconomic distribution of the aggregate interactions of users with $D_s(u)=0$, for all classes from class 1 (b) to class 9 (r). }
  \label{figs15}
\end{figure}

\newpage

\begin{figure}[ht!]
  \centering
  \includegraphics[width=\linewidth]{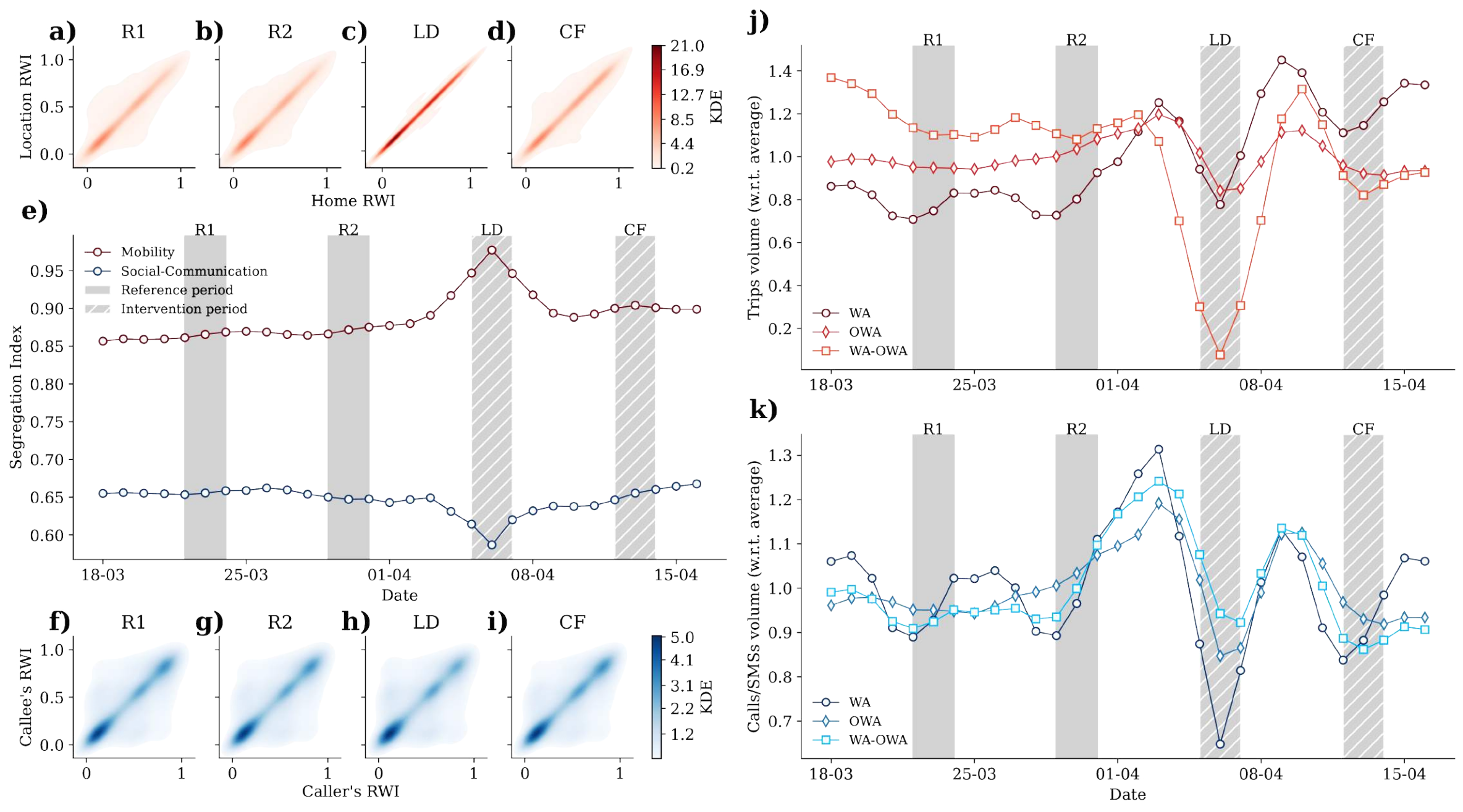}
  \caption{\textbf{Spatial effects on global segregation.} Same calculations of Fig. 1 in the main text, here made without removing links between nodes located at the same location (respectively, movements to the home locations and calls/SMSs to users with the same home location as the caller). (a-d) SE assortativity matrices (shown as the kernel density of the joint probability of RWIs) of the mobility network during the two reference (R1 and R2), lockdown (LD), and curfew (CF) periods. (e) The dynamics of the $\rho$ SE assortativity index computed for the mobility (red) and social communication (blue) networks. (f-i) Same as (a-d) but for the social communication network. (j) Relative number of travels within WA, OWA, and between the areas WA-OWA (also accounting for OWA-WA trips). (k) Number of communication events between people living in WA and OWA, or between the two geographic areas. All curves are normalized by their average computed over the full data period. For calculations on panels (e), (j), and (k), we used 3-day symmetric rolling time windows with a 1-d shift to obtain aggregated networks around the middle day at time $t$ of the actual window.}
  
  \label{figs2}
\end{figure}

\newpage

\begin{figure}[ht!]
  \centering
  \includegraphics[width=\linewidth]{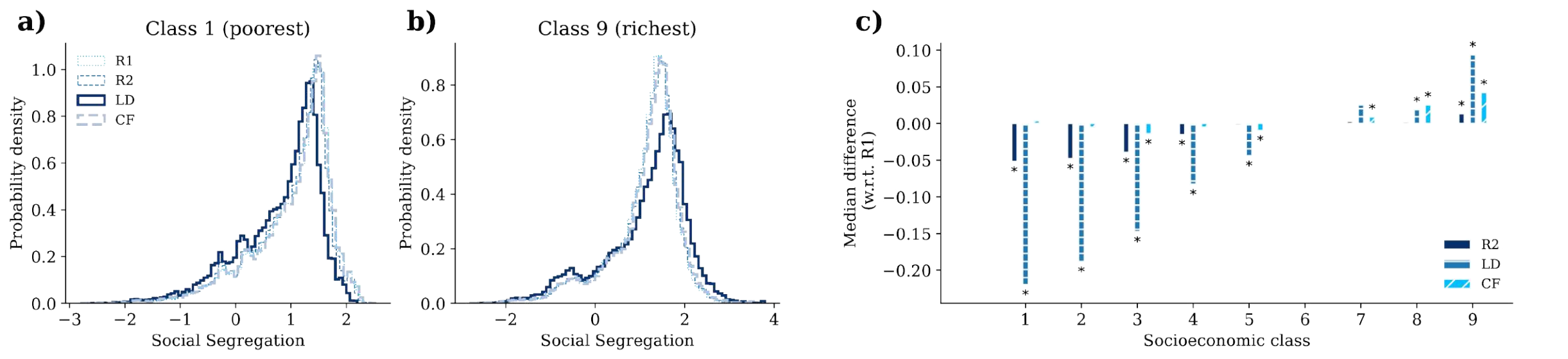}
  \caption{\textbf{Spatial effects on individual segregation in the social communication network.} Same calculations of the bottom panels of Fig. 2 in the main text, here made without removing links between nodes located at the same location (calls/SMSs to users with the same home location as the caller). The $P(r_u(t))$ individual assortativity index distributions for the poorest (class 1 in a) and the richest (class 9 in b) SE classes for the two reference periods (R1 and R2, thin dashed lines), the lockdown (LD, solid line), and curfew (CF, dashed thick line). Panel (c) depicts the pairwise differences of median assortativity values of $P(r_u(t))$ for each of the nine SE groups in the social communication network. Differences are calculated pairwise between R1 and the R2, LD, and CF periods. The asterisk symbols over the bars (when bars are positive, otherwise under them) in panels (c) indicate statistically significant differences computed with the one-tailed Mann-Withney U-test (with $\mbox{p-value} < 0.01$) }
  \label{figs3}
\end{figure}

\newpage

\begin{figure}[ht!]
  \centering
  \includegraphics[width=0.6\linewidth]{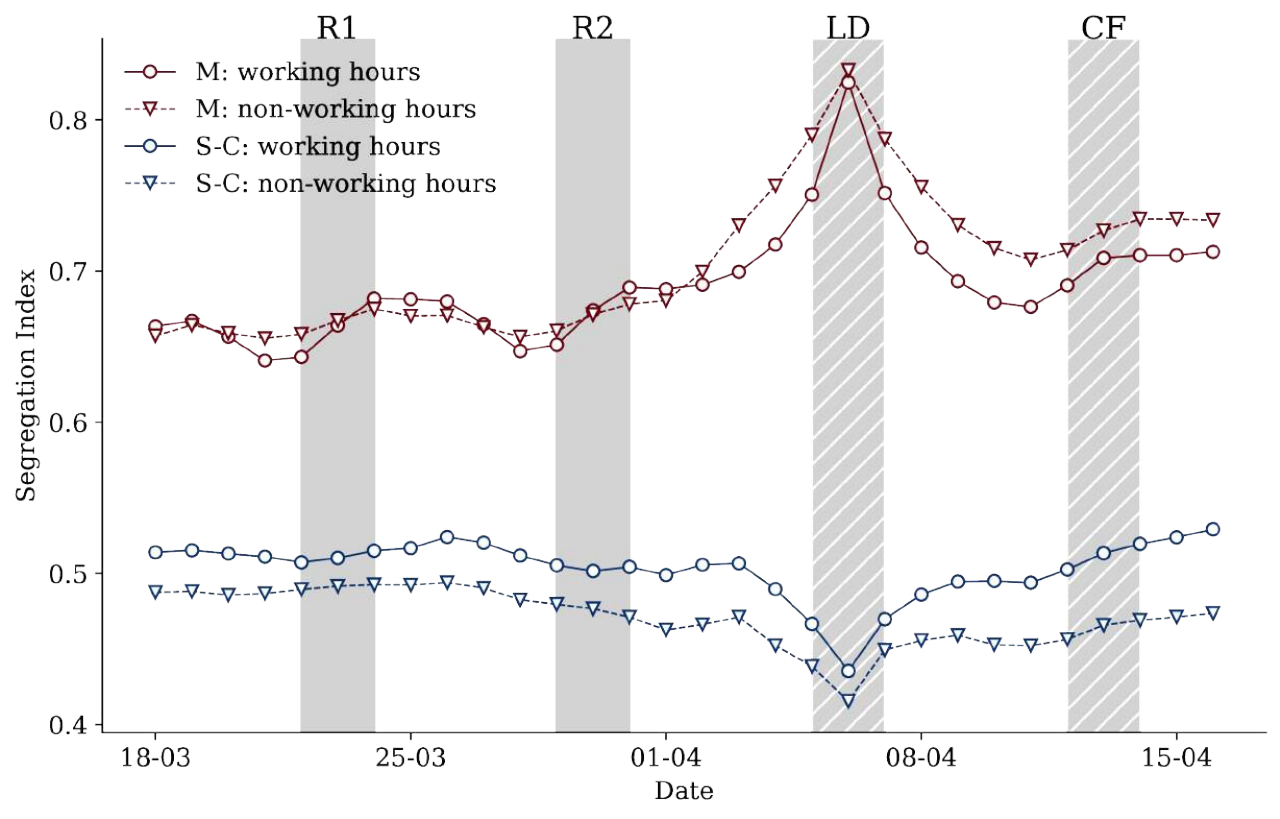}
  \caption{\textbf{Effects of professional activities.} Segregation curves for mobility (red) and social communication (blue) networks, obtained by explicitly separating the activities during office hours (9 AM - 7 PM, solid lines) from out-of-office hours (dashed lines).}
  \label{figs4}
\end{figure}

\newpage

\begin{figure}[ht!]
    \centering
  \includegraphics[width=0.8\linewidth]{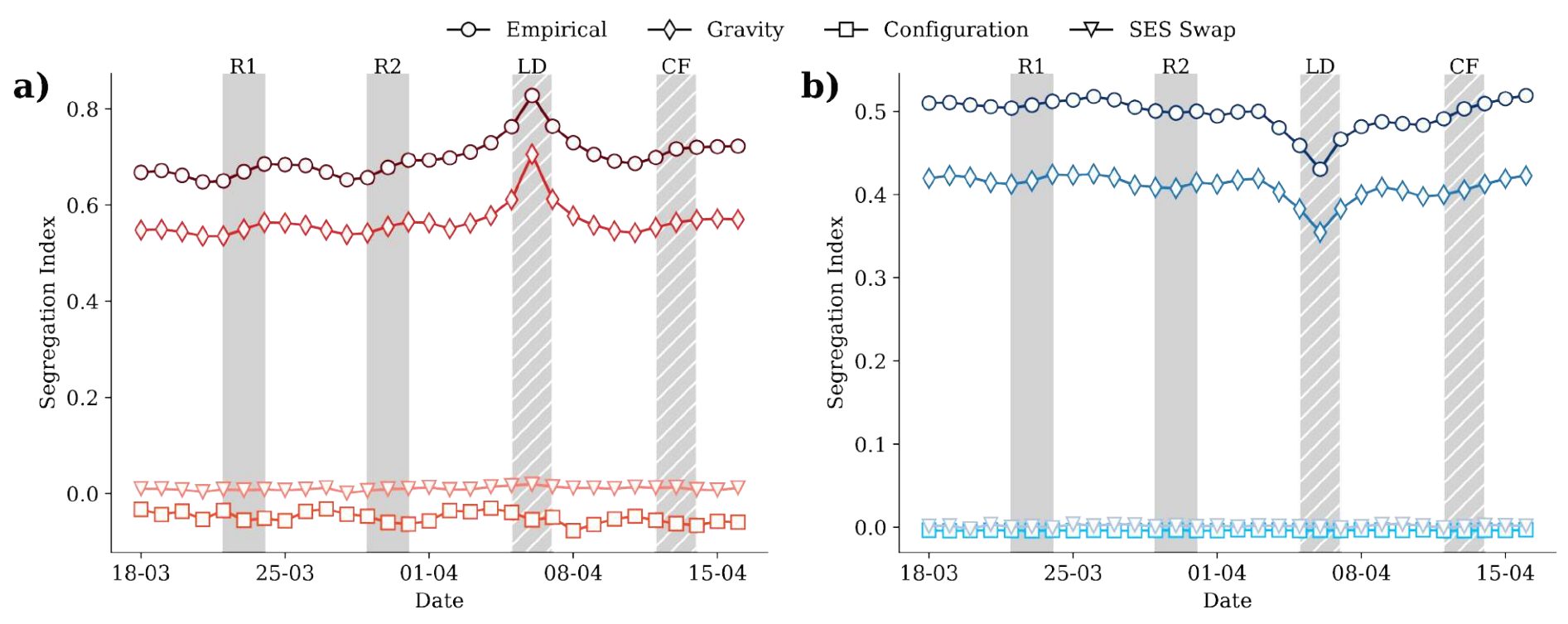}
  \caption{\textbf{Reference models.} Evolution of segregation levels obtained from the data (darkest curves with circles, the same as in Fig. 1E in the main text) and from reference models (lighter curves) in the (a) mobility and (b) social communication networks. As reference models the gravity model (diamond curve), the configuration model (square curve), and SES label swapped reference model (triangle curve) were considered.}
  \label{figs5}
\end{figure}

\newpage

\begin{figure}[ht!]
    \centering
  \includegraphics[width=0.8\linewidth]{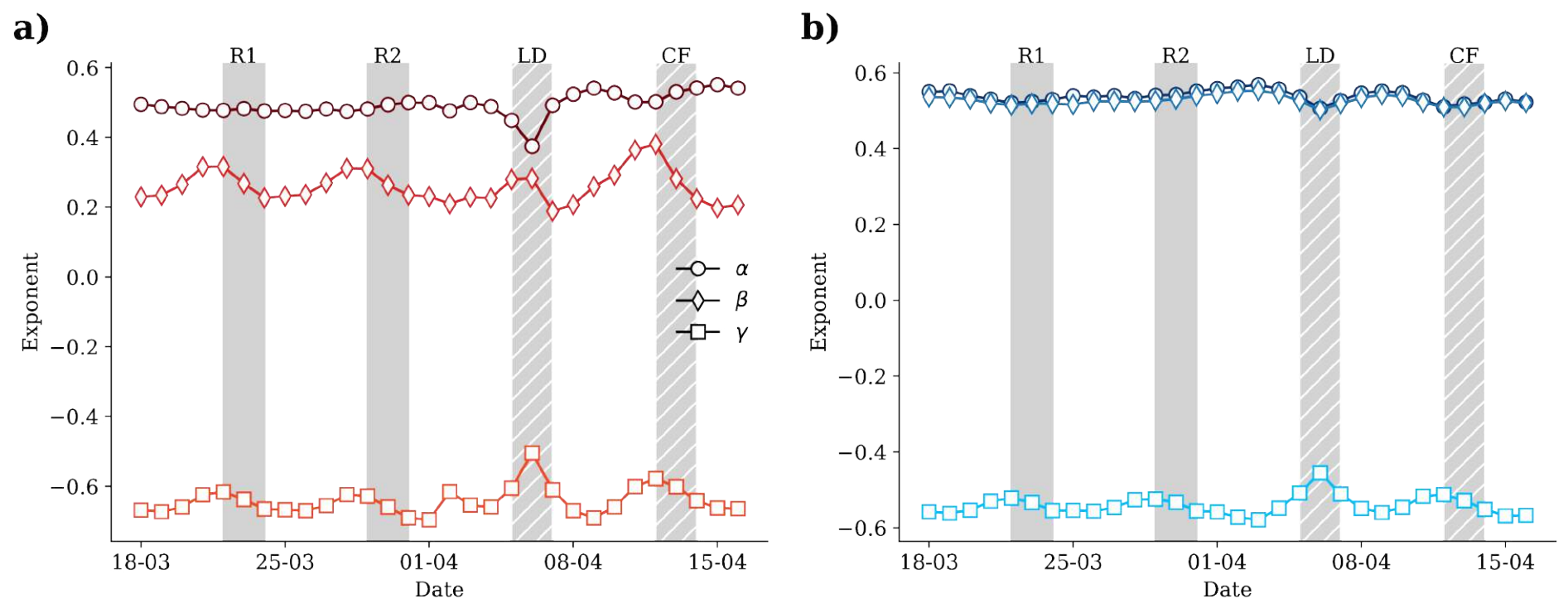}
  \caption{\textbf{Gravity models} The three exponents (alpha, beta, gamma) of the gravity model, fitted with an OLS linear model, in the (a) mobility and (b) social communication network.}
  \label{figs6}
\end{figure}

\newpage

\begin{figure}[ht!]
    \centering
  \includegraphics[width=1\linewidth]{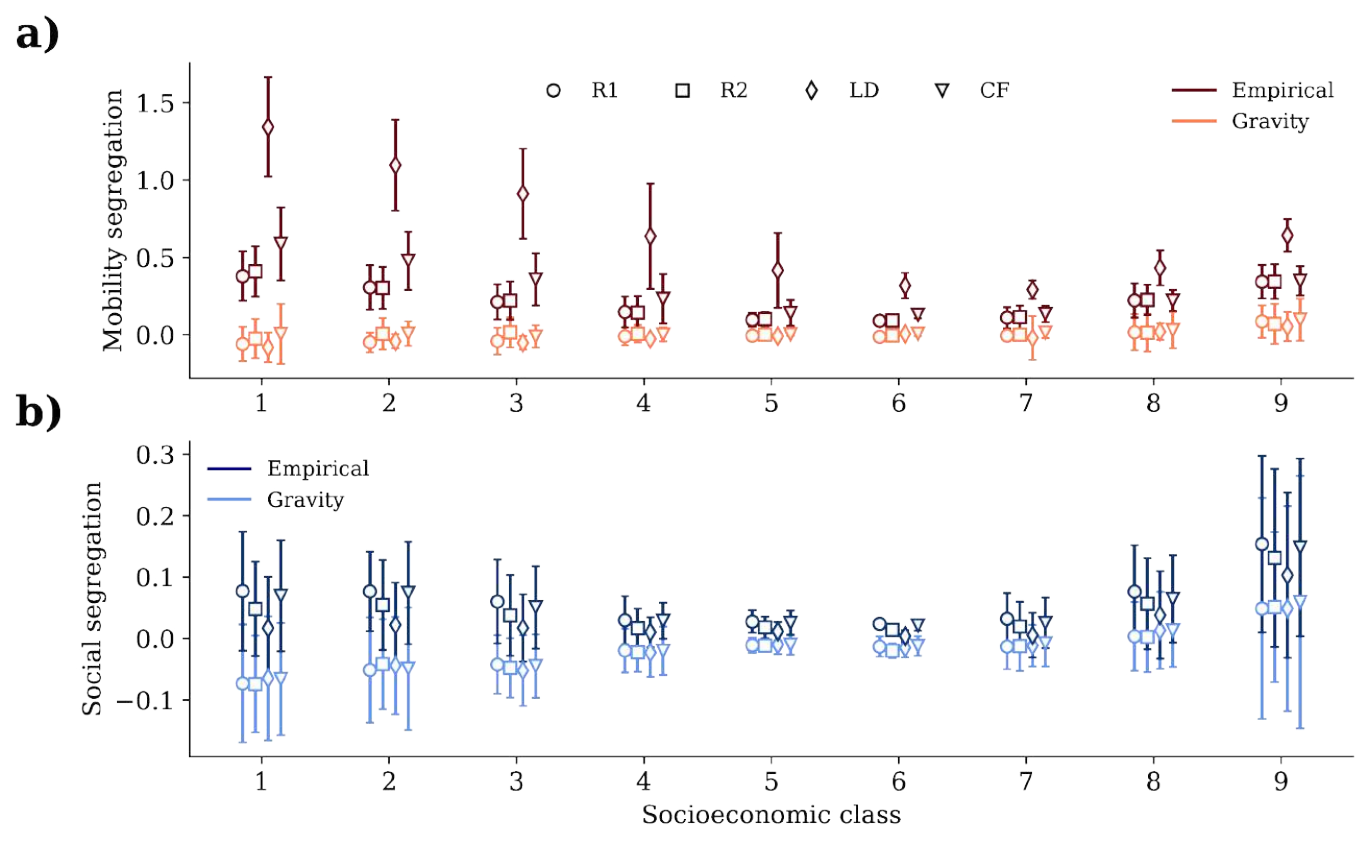}
  \caption{\textbf{Individual gravity model.} (a) Mean individual segregation (with standard deviation) for all socioeconomic classes in mobility networks, from empirical data (dark red) and from the gravity model (light red), for R1 (circles), R2 (squares), LD (diamonds), and CF (triangles). (b) Mean individual segregation (with standard deviation) for all socioeconomic classes in social communication networks aggregated at the location level, from empirical data (blue) and from the gravity model (light blue), for R1 (circles), R2 (squares), LD (diamonds), and CF (triangles).
  }
  \label{figs16}
\end{figure}

\newpage

\begin{figure}[ht!]
    \centering
  \includegraphics[width=1\linewidth]{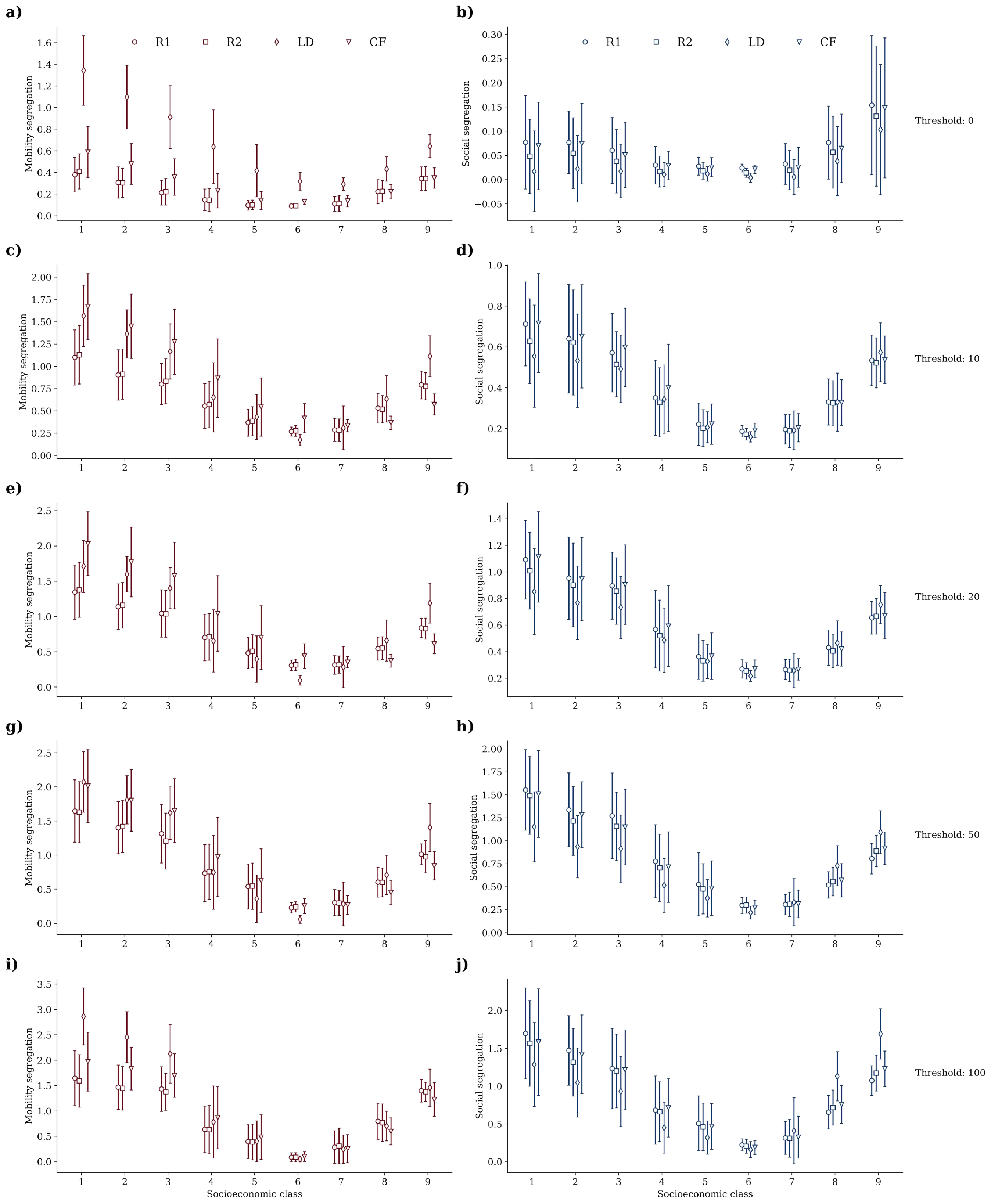}
  \caption{\textbf{Weight thresholding.} Mean individual segregation (with standard deviation) for all socioeconomic classes in mobility (left panels) and social communication (right panels) networks, for R1 (circles), R2 (squares), LD (diamonds), and CF (triangles), and for different weight threshold values (both figures on a single row refer to the threshold value on the right of the row).
  }
  \label{figs17}
\end{figure}

\newpage
\clearpage

\begin{table}[ht!]
\centering
\caption{School closures}
\begin{tabular}{c|l} 
0 & No measures \\
1 & Recommend closing \\
2 & Require closing (only some levels or categories, e.g. just high school, or just public schools) \\
3 & Require closing all levels
\end{tabular}
\label{tabs2}
\end{table}

\begin{table}[ht!]\centering
\caption{Workplace closures}
\begin{tabular}{c|l} 
0 & No measures \\
1 & Recommend closing (or work from home)\\
2 & Require closing (or work from home) for some sectors or categories of workers\\
3 & Require closing (or work from home) all but essential workplaces (e.g. grocery stores, doctors)\\
\end{tabular}
\label{tabs3}
\end{table}

\begin{table}[ht!]\centering
\caption{Cancel public events}
\begin{tabular}{c|l} 
0& No measures\\
1 & Recommend cancelling\\
2 & Require cancelling\\
\end{tabular}
\label{tabs4}
\end{table}

\begin{table}[ht!]\centering
\caption{Restrictions on gatherings}
\begin{tabular}{c|l} 
0 & No restrictions\\
1 & Restrictions on very large gatherings (the limit is above 1,000 people)\\
2 & Restrictions on gatherings between 100-1,000 people\\
3 & Restrictions on gatherings between 10-100 people\\
4 & Restrictions on gatherings of less than 10 people\\
\end{tabular}
\label{tabs5}
\end{table}

\begin{table}[ht!]\centering
\caption{Close public transport}
\begin{tabular}{c|l} 
0 & No measures\\
1 & Recommend closing (or significantly reduce volume/route/means of transport available)\\
2 & Require closing (or prohibit most citizens from using it)\\
\end{tabular}
\label{tabs6}
\end{table}

\begin{table}[ht!]\centering
\caption{Public information campaigns}
\begin{tabular}{c|l} 
0 & No COVID-19 public information campaign\\
1 &  public officials urging caution about COVID-19\\
2 &  coordinated public information campaign (e.g. across traditional and social media)\\
\end{tabular}
\label{tabs7}
\end{table}

\begin{table}[ht!]\centering
\caption{Stay at home}
\begin{tabular}{c|l} 
0 & No measures\\
1 & recommend not leaving house\\
2 & require not leaving house with exceptions for daily exercise, grocery shopping, and ‘essential’ trips\\
3 & Require not leaving house with minimal exceptions (e.g. allowed to leave only once every few days, or only one person can leave at a time, etc.)\\
\end{tabular}
\label{tabs8}
\end{table}

\begin{table}[ht!]\centering
\caption{Restrictions on internal movement}
\begin{tabular}{c|l} 
0 & No measures\\
1 & Recommend movement restriction\\
2 & Restrict movement\\
\end{tabular}
\label{tabs9}
\end{table}

\begin{table}[ht!]\centering
\caption{International travel controls}
\begin{tabular}{c|l} 
0 & No measures\\
1 & Screening\\
2 & Quarantine arrivals from high-risk regions\\
3 & Ban on high-risk regions\\
4 & Total border closure\\
\end{tabular}
\label{tabs10}
\end{table}

\begin{table}[ht!]\centering
\caption{Median values and standard deviations of individual assortativity index distributions. Values are computed for nodes (locations in the mobility network or people in the social network) from the nine SE classes during the two reference periods (R1 and R2) and the intervention periods (LD and CF). Standard deviation values are shown in parentheses.}

\small
\resizebox{\textwidth}{!}{
\centering

\begin{tabular}{|c|ccccccccc|} 
 \hline
 $G_M$ & class 1 & class 2 & class 3 & class 4 & class 5 & class 6 & class 7 & class 8 & class 9 \\ [0.5ex] 
 \hline
R1 & 0.41 (0.16) & 0.33 (0.14) & 0.24 (0.11) & 0.16 (0.1) & 0.1 (0.04) & 0.09 (0.01) & 0.08 (0.07) & 0.2 (0.11) & 0.34 (0.11) \\ 
R2 & 0.4 (0.16) & 0.33 (0.14) & 0.24 (0.12) & 0.15 (0.1) & 0.1 (0.04) & 0.1 (0.01) & 0.09 (0.07) & 0.21 (0.1) & 0.32 (0.11)\\ 
LD & 1.28 (0.32) & 1.1 (0.3) & 0.92 (0.29) & 0.62 (0.33) & 0.48 (0.24) & 0.31 (0.08) & 0.28 (0.06) & 0.46 (0.11) & 0.62 (0.11) \\ 
CF & 0.59 (0.24) & 0.47 (0.19) & 0.4 (0.17) & 0.24 (0.16) & 0.15 (0.08) & 0.13 (0.02) & 0.12 (0.05) & 0.23 (0.07) & 0.34 (0.09)\\ \hline
$G_S$ &  &  &  &  &  &  &  &  &  \\ [0.5ex] 
 \hline
R1 & 1.24 (0.71) & 0.97 (0.62) & 0.78 (0.52) & 0.39 (0.32) & 0.17 (0.18) & 0.12 (0.17) & 0.18 (0.34) & 0.86 (0.61) & 1.28 (0.75) \\ 
R2 & 1.18 (0.71) & 0.92 (0.6) & 0.75 (0.51) & 0.38 (0.31) & 0.16 (0.17) & 0.12 (0.18) & 0.18 (0.34) & 0.87 (0.62) & 1.3 (0.77) \\ 
LD & 1.02 (0.73) & 0.78 (0.62) & 0.64 (0.52) & 0.31 (0.3) & 0.12 (0.17) & 0.12 (0.21) & 0.21 (0.4) & 0.88 (0.73) & 1.38 (0.93) \\ 
CF & 1.24 (0.72) & 0.96 (0.62) & 0.77 (0.52) & 0.39 (0.32) & 0.16 (0.17) & 0.12 (0.18) & 0.19 (0.35) & 0.89 (0.63) & 1.32 (0.78) \\ 
 \hline
\end{tabular}
}
\label{tabs1}
\end{table}

\begin{table}[ht!]\centering
\caption{List of p-values (order of magnitude) obtained from the statistical comparison of distributions (shown in Fig. 2 in the main text) with the one-tailed Mann-Withney U-test.}

\small
\resizebox{\textwidth}{!}{
\centering

\begin{tabular}{|c|ccccccccc|}
 \hline
 $G_M$ & class 1 & class 2 & class 3 & class 4 & class 5 & class 6 & class 7 & class 8 & class 9 \\ [0.5ex] 
 \hline
R2-R1 & $10^{-1}$ & $10^{-1}$ & $10^{-1}$ & $10^{-1}$ & $10^{-1}$ & $10^{-1}$ & $10^{-1}$ & $10^{-1}$ & $10^{-1}$ \\ 
LD-R1 & $10^{-20}$ & $10^{-21}$ & $10^{-19}$ & $10^{-9}$ & $10^{-9}$ & $10^{-10}$ & $10^{-8}$ & $10^{-10}$ & $10^{-10}$ \\ 
CF-R1 & $10^{-7}$ & $10^{-7}$ & $10^{-7}$ & $10^{-3}$ & $10^{-3}$ & $10^{-8}$ & $10^{-2}$ & $10^{-1}$ & $10^{-1}$ \\ \hline
$G_S$ &  &  &  &  &  &  &  &  &  \\ [0.5ex] 
 \hline
R2-R1 & $10^{-15}$ & $10^{-15}$ & $10^{-11}$ & $10^{-6}$ & $10^{-4}$ & $10^{-1}$ & $10^{-1}$ & $10^{-1}$ &  $10^{-4}$ \\ 
LD-R1 & $10^{-224}$ & $10^{-210}$ & $10^{-169}$ & $10^{-137}$ & $10^{-205}$ & $10^{-101}$ & $10^{-1}$ & $10^{-13}$ & $10^{-77}$  \\ 
CF-R1 & $10^{-1}$ & $10^{-3}$ & $10^{-4}$ & $10^{-2}$ & $10^{-4}$ & $10^{-5}$ & $10^{-1}$ & $10^{-5}$ & $10^{-16}$   \\ 
 \hline
\end{tabular}
}
\label{tabs11}
\end{table}

\clearpage
%Bibliography
%\bibliographystyle{unsrt}  

%--/Paper--

\end{document}